\documentclass[num-refs]{ametsocV6.1}

\newcommand\BibTeX{{\rmfamily B\kern-.05em \textsc{i\kern-.025em b}\kern-.08em
	T\kern-.1667em\lower.7ex\hbox{E}\kern-.125emX}}

\usepackage{moreverb}

\usepackage[utf8]{inputenc}
\usepackage[T1]{fontenc}
\usepackage{amsmath}
\usepackage{esint}
\usepackage{amssymb}
\usepackage{graphicx}
\usepackage{placeins}
\usepackage{graphicx}
\usepackage{tikz,pgfplots}
\usetikzlibrary{positioning}
\usepackage[caption=false]{subfig}
\usepackage{float}
\usepackage{subeqnarray}
\usepackage{lipsum}
\usepackage{siunitx}
\usepackage{color}
\usepackage[T1]{fontenc}
\usepackage{epic}

\usepackage{eurosym}
\usepackage[colorinlistoftodos]{todonotes}
\usepackage{comment}

\newcommand{\be}{\begin{equation}}
\newcommand{\ee}{\end{equation}}
\newcommand{\bae}{\begin{eqnarray}}
\newcommand{\eae}{\end{eqnarray}}
\newcommand{\bse}{\begin{subeqnarray}}
\newcommand{\ese}{\end{subeqnarray}}

\title{Feedback Processes causing an AMOC Collapse in the Community Earth System Model}
\clearpage
\authors{Elian Vanderborght\aff{a}\correspondingauthor{Elian Vanderborght, e.y.p.vanderborght@uu.nl}, 
René M. van Westen \aff{a} and
Henk A. Dijkstra \aff{a,b} }

\affiliation{\aff{a}Institute for Marine and Atmospheric research Utrecht, Department of Physics, 
Utrecht University, Utrecht, the Netherlands\\
\aff{b} Centre for Complex Systems Studies,  Utrecht University, Utrecht, the Netherlands. }

\abstract{The Atlantic Meridional Overturning Circulation (AMOC) is recognized as a  tipping element within the global climate system. Central to its tipping behavior is the salt-advection feedback mechanism, which has been extensively studied in box models and models of intermediate complexity. However, in contemporary, highly complex climate models, the importance and functioning of this feedback mechanism is less clear due to the intricate interplay of numerous ocean-atmosphere-sea ice feedbacks. In this study, we conduct a detailed mechanistic analysis of an AMOC collapse under quasi-equilibrium forcing conditions using the Community Earth System Model (CESM). By reconstructing the AMOC strength from  the meridional density contrast across the Atlantic Ocean, we demonstrate that AMOC stability can be related to  the Atlantic freshwater budget, revealing several  important feedbacks. The dominant contribution is the destabilising salt-advection feedback, which is quantified through a negative sign of the overturning freshwater transport at 34$^{\circ}$S, indicated by $F_{\mathrm{ovS}}$.  Other feedbacks  are related to changes in North Atlantic sea-ice melt (destabilising), ocean-atmosphere freshwater fluxes (destabilising) and gyre transports (stabilising).  Our study clarifies the role of $F_{\mathrm{ovS}}$ as an indicator of the background state stability of the AMOC. As many modern climate models have a positive $F_{\mathrm{ovS}}$ bias this implies that their AMOC is too stable which leads to an underestimation of the risk of an AMOC collapse under climate change.}

\begin{document}		
\maketitle

\statement{The Atlantic Meridional Overturning Circulation (AMOC) is a key component and tipping element in the climate system.
	Recent climate model simulations demonstrated that the AMOC can tip under the input of freshwater in the Atlantic Ocean.
	A potential AMOC tipping event has severe climate and societal impacts and it is therefore important 
	to understand the feedback mechanisms that cause the AMOC tipping event.  
	This study identifies and quantifies these feedback mechanisms and their effects on AMOC stability. 
	Our findings show that the dominant salt-advection feedback is responsible for AMOC tipping
	and is only activated   when the AMOC carries net salinity into the Atlantic Ocean. This finding establishes 
	a physical connection between the Atlantic salt transport and the salt-advection feedback. Understanding 
	this connection is important for assessing the risk of AMOC tipping within the 21$^\text{st}$ century.}
\newpage

\section{Introduction}{\label{Introduction}}

The Atlantic Meridional Overturning Circulation (AMOC) is a part of the global ocean circulation that effectively transports heat, salt and carbon through the Atlantic Ocean \citep{johns2011continuous,broecker1991great}. 
The meridional heat transport by the AMOC plays a significant role in maintaining the temperate climate of Northwestern Europe \citep{palter2015role}.
Variations in AMOC strength are also linked to the latitudinal position of the Inter-Tropical Convergence Zone \citep{schneider2014migrations} and the Arctic sea-ice extent \citep{yeager2017recent}. 
Given this important role of the AMOC in the climate system \citep{rahmstorf2002ocean, clark2002role, zhang2019review},
the AMOC is closely monitored along the RAPID-MOCHA measurement array at 26$^\circ$N since 2004 \citep{srokosz2015observing}. 
and  shows a time mean (2004 -- 2022)  strength of 16.9$\pm$4.6~Sv (1~Sv $\equiv$ 10$^\text{6}$ m$^\text{3}$s$^{\text{-1}}$). The AMOC 
is likely to weaken under anthropogenically forced climate change in the remainder of the 21$^{\mathrm{st}}$ century \citep{jackson2020impact,weijer2020cmip6}. 

The AMOC has been identified as a key tipping element in the climate system \citep{lenton2008tipping, armstrong2022exceeding}. 
This is supported by paleo-climate evidence, showing abrupt climate transitions linked to strong AMOC variations \citep{dansgaard1993evidence,de2003millennial}. 
\cite{stommel1961thermohaline} showed that a highly idealized 2-box model of the AMOC has two stable states under 
identical freshwater forcing conditions.  The emergence of this multi-stable AMOC regime originates from the salt-advection feedback 
This feedback is characterized by a (North Atlantic) freshwater perturbation that weakens the AMOC, leading to a reduced northward salinity transport and thereby amplifying the initial perturbation \citep{Marotzke2000}. 

\cite{rahmstorf1996freshwater} linked the strength of the salt-advection feedback to the freshwater transport carried by the AMOC at the southern boundary (34$^\circ$S) of the Atlantic Ocean, indicated here by $F_{\mathrm{ovS}}$. 
This was done in an extended version of the \cite{rooth1982hydrology} box model.
When $F_{\mathrm{ovS}}$ is positive (negative), the AMOC carries net freshwater (salinity) into the Atlantic Ocean.
\cite{huisman2010indicator} provided a physical link between the $F_{\mathrm{ovS}}$ and the salt-advection feedback by constructing an integrated freshwater balance 
for the Atlantic Ocean and showed that Atlantic freshwater perturbations were only amplified for negative $F_{\mathrm{ovS}}$. 
This argument provided strong support for $F_{\mathrm{ovS}}$ as an indicator for a multi-stable AMOC, 
where a negative $F_{\mathrm{ovS}}$ indicates the existence of a stable collapsed AMOC state under identical freshwater forcing \citep{sijp2012characterising}.   

Building on this theoretical framework, numerous studies employing models of varying complexity confirmed the 
significance of $F_{\mathrm{ovS}}$ as an AMOC stability indicator \citep{de2005atlantic, dijkstra2007characterization, hawkins2011bistability, jackson2013shutdown, van2024physics}. 
Hydrographic observations show that the present-day $F_{\mathrm{ovS}}$ is negative \citep{bryden2011south,garzoli2013south, arumi2024multi} 
which  is in stark contrast with the persistent positive $F_{\mathrm{ovS}}$ biases found in the latest generations of climate models \citep{mecking2017effect,van2024physics}. 
The implication would be  that climate models underestimate the risk of AMOC tipping under climate change \citep{jackson2013shutdown,liu2017overlooked}. This results in (unrealistically) high tipping thresholds (e.g., 
enhanced Greenland Ice Sheet melt and global warming)   which  are likely not reached within the  
21$^{\mathrm{st}}$ century \citep{lenton2008tipping,armstrong2022exceeding}.  

Despite the extensive research underscoring the importance of $F_{\mathrm{ovS}}$,  key concerns remain regarding its role in AMOC 
stability \citep{gent2018commentary}. First, in a steady-state, the Atlantic freshwater budget represents a balance between the surface freshwater flux and the total horizontal freshwater transport. When a freshwater perturbation is introduced in the North Atlantic, this balance is disrupted, and the system evolves toward a new equilibrium.  In the case of overturning  freshwater transport, this adjustment is influenced by the salt-advection feedback.  However, model studies indicate that  other feedbacks, such as those due to the gyre freshwater transport, can sometimes 
exceed the salt-advection feedback \citep{mecking2016stable,jackson2013shutdown} which  can affect AMOC stability \citep{cimatoribus2014meridional}.  Thus, the importance of the salt-advection feedback depends on the relative influence 
of other feedback processes within the Atlantic freshwater budget \citep{gent2018commentary,weijer2019stability}. 
The complex interplay of multiple feedbacks in coupled climate models further complicates the interpretation of 
$F_{\mathrm{ovS}}$ as a stability indicator.

Second, while arguments support the physical relevance of $F_{\mathrm{ovS}}$ within the context of the total Atlantic freshwater balance, 
there is ambiguity in the relation between the Atlantic freshwater balance and AMOC stability in modern climate models \citep{mecking2016stable, weijer2019stability}. 
An alternative approach to the total Atlantic freshwater balance is by analysing the meridional density (or buoyancy) contrast between the North Atlantic and more southern latitudes.
The AMOC responses were successfully reconstructed from thermal wind balance when using this meridional density contrast \citep{butler2016reconstructing, bonan2022transient, haskins2019explaining, haskins2020temperature, jansen2018transient},
underlining the importance of the meridional density gradient rather than the total freshwater balance. In this approach, several  studies
showed  that the meridional density gradient was not affected by South Atlantic freshwater transport anomalies 
\citep{haines2022variability, cheng2013atlantic, mignac2019decoupled} challenging the relation between $F_{\mathrm{ovS}}$ 
and the AMOC strength. 

The aim of this study is to provide a mechanistic and quantitive 
description of an AMOC collapse in a modern complex climate model.  This mechanistic description addresses the 
aforementioned points by quantifying the different AMOC feedback strengths and unravels the role of $F_{\mathrm{ovS}}$ 
in AMOC stability.  Thereto we will analyse the recent  quasi-equilibrium hosing simulation performed with the Community 
Earth System Model (CESM) \citep{van2023asymmetry, van2024physics}, which shows an AMOC collapse 
under the slowly increasing freshwater flux forcing. The quasi-equilibrium nature of this simulation provides an 
ideal case for investigating AMOC feedback mechanisms, enabling us to demonstrate how these feedbacks 
strengthen or weaken under varying hosing intensities. Our study is organized as follows: Section~\ref{Methods} 
briefly describes the CESM setup, and introduces the AMOC diagnostics used.  Section~\ref{Results} presents the 
results using this diagnostic, while Section~\ref{Anl} provides the analysis and interpretation of the findings. Finally, 
Section~\ref{SD} provides a summary and discussion. 

\section{Methods}\label{Methods}
\subsection{Model}\label{Model}

We analyse the  simulation presented in 
\cite{van2024physics}, performed with the CESM (version 1.0.5. in the f19 g16 configuration).
The CESM has a 1$^\circ$ ocean (Parallel Ocean Program version 2, POP2 \citep{smith2010parallel}), a $1^\circ$ sea-ice model (Community Ice Code version 4, CICE4 \citep{hunke2010cice}), 
and a 2$^\circ$ atmosphere/land component (Community Atmosphere Model version 4, CAM4 \citep{neale2013mean}). 
Upon initialization from a spin-up solution  \citep{baatsen2020middle}, the CESM is well equilibrated and  the pre-industrial 
greenhouse forcing conditions are kept constant. The standard monthly-averaged output is converted to yearly averages.

The hosing (i.e., freshwater flux into the ocean) is applied over the Atlantic Ocean between 20$^\circ$N and 50$^\circ$N. 
To conserve salinity, the applied surface freshwater flux is compensated globally elsewhere.
The freshwater flux forcing, $F_H$, linearly increases at a rate of 3$\times$10$^{-4}$~Sv~yr$^{-1}$, similar to \cite{hu2012role}. 
This is a slow hosing rate ensuring  quasi-equilibrium conditions, i.e.,  the CESM stays relatively close to its (statistical) equilibria  \citep{van2024role}. 

\subsection{Reconstructing the AMOC strength from the buoyancy field}\label{Data}

The meridional overturning streamfunction is given by:
\begin{equation}
	\psi(t,y,z)=-\int_{-H}^z \left[ \int^{x_E}_{x_W} v(t,x,y,z') dx \right] dz'
	\label{MStrF}
\end{equation}
where $v$ is the meridional velocity, $H$ is the water column depth and $x_E$ and $x_W$ refer to the western 
and eastern boundary of the Atlantic Ocean, respectively. Following \cite{butler2016reconstructing}, the interior overturning streamfunction, $\psi_{\mathrm{int}}(t, z)$, is defined as $ \max \{\psi(t,y,z) \mid y \in [0^\circ \text{N}, 30^\circ \text{N}]\}$.

To analyse the different AMOC feedbacks, we aim to reconstruct the AMOC strength using  thermal wind balance (TWB). 
Specifically, the zonally-integrated meridional velocity $V$,  is in TWB 
with the zonal basin-scale potential density difference $\Delta_x \rho$ (east minus west) \citep{cessi2009eddy}. This relation can be 
expressed as: 
\begin{equation} 
	\frac{\partial V}{\partial z} = \frac{g}{\rho_0 f}\Delta_x \rho, \label{Basin_Scale_Velocity} 
\end{equation}
where $g$ is the gravitational acceleration (= 9.81 m s$^{-2}$), $\rho_0$ is a reference density (= 1025 kg m$^{-3}$), 
$f$ is the Coriolis parameter.   Physical arguments from \cite{marotzke1997boundary} and \cite{kuhlbrodt2007driving} 
explain how  $\Delta_x \rho$, translates into a basin-scale meridional potential  density gradient.  Specifically, this 
relation is given by $\Delta_x \rho = C\Delta_y \rho$, where the dimensionless  proportionality constant $C$ incorporates 
the effects of basin geometry and boundary layer structure \citep{gnanadesikan1999simple}.

By combining Equations (\ref{MStrF}) and (\ref{Basin_Scale_Velocity}), we evaluate the TWB overturning transport as:
\begin{equation}
	\frac{\partial^2 \hat{\psi}_{\mathrm{int}}}{\partial z^2} = \frac{g C}{\rho_0 f} \Delta_y \rho (z),
	\label{TWB}
\end{equation}
where the hat superscript refers to the TWB assumption. 
Equation (\ref{TWB}) is solved with the boundary conditions $\hat{\psi}_\mathrm{int}(t, 0) = 0$ Sv at the surface 
and $\hat{\psi}_{\mathrm{int}}(t, -H) = 0$ Sv at the bottom of the ocean.
In this way, we reconstruct $\psi_{\mathrm{int}}(t,z)$, as simulated in the CESM, from the buoyancy field by assuming that the AMOC is in TWB 
\citep{nikurashin2012theory}.  We define $\Psi_{\mathrm{int}}(t) = \psi_{\mathrm{int}}(t, z_{int})$  (or equivalently  $\hat{\Psi}_{\mathrm{int}}(t) = \hat{\psi}_{\mathrm{int}}(t, z_{int})$) as the (reconstructed) interior AMOC strength, taking $z_{int} = -1000$ m because $\psi_{\mathrm{int}}(t, z)$ has a 
maximum around $1,000$~m depth. 

The buoyancy reconstructed interior streamfunction $\hat{\psi}_\mathrm{int}$ and AMOC strength 
$\hat{\Psi}_{\mathrm{int}}$ depend 
on both the shape of the density profile and the magnitude of the density contrast \citep{butler2016reconstructing}. 
Hence, we decompose the density contrast as $\Delta_y \rho(z) = \langle \Delta_y \rho \rangle + \gamma(z)$, 
where $\langle \Delta_y \rho \rangle$ is the vertically-averaged meridional potential density difference, and 
$\gamma(z)$ represents the deviations from this vertical average.  Given this decomposition, changes in the 
reconstructed AMOC strength are given as:
\begin{equation}
	\delta \hat{\Psi}_{\mathrm{int}} = \delta \{\hat{\Psi}_{\mathrm{int}}\}_{\langle \Delta_y \rho \rangle}+\delta \{\hat{\Psi}_{\mathrm{int}}\}_{\gamma},
	\label{PerturbationPsi} 
\end{equation}
where $\delta \hat{\Psi}_{\mathrm{int}} = \hat{\Psi}_{\mathrm{int}}(t_2) - \hat{\Psi}_{\mathrm{int}}(t_1)$, with $t_1$ and $t_2$ 
denoting two consecutive time instances.  The notation $\delta \{F\}_G$ denotes the change in $F$ resulting 
from a change in $G$. We later compute the cumulative change in $F$, resulting from a change in $G$, as:
\begin{equation} 
	\{F\}_{G}(t) = \sum^{t \leq t_i}_{t_i=0}\delta \{F\}_{G}(t_i),
\end{equation} 
where $t_i$ denotes the discrete time instances in the model simulation.

The first term on the right hand side of (\ref{PerturbationPsi}) represents the change in $\Psi_{\mathrm{int}}$ 
due to variations in the  vertically-averaged potential density contrast, which can be expressed as: 
\begin{equation}
	\delta \{\hat{\Psi}_{\mathrm{int}}\}_{\langle \Delta_y \rho \rangle}=\frac{gC}{\rho_0 f}D^2\delta\langle \Delta_y \rho \rangle,
\end{equation}
where $D^2 = 1/2 \left(-z_{\mathrm{int}}H-z_{\mathrm{int}}^2\right)$,  reflecting 
the chosen depth for $\hat{\Psi}_{\mathrm{int}}$.  The expression for $D^2$ is found when solving equation (\ref{TWB}) with the 
appropriate boundary conditions, and with $H\approx$ 4,700~m we find $D\approx$ 1,300~m. 
The second term on the right hand side in Equation (\ref{PerturbationPsi}), $\delta \{\hat{\Psi}_{\mathrm{int}}\}_{\gamma}$, accounts for changes in $\hat{\Psi}_{\mathrm{int}}$ due to modifications in the vertical structure of the meridional density contrast profile and consequently quantifies the effect of changes in stratification on the interior AMOC strength. 

The vertically-averaged meridional potential density differences (Equation~\ref{PerturbationPsi}) can be further 
decomposed into a salinity ($S$) and potential temperature ($T$) contribution.
Specifically, employing a linear equation of state we write:
\begin{equation}
	\delta\{\hat{\Psi}_{\mathrm{int}}\}_{\langle \Delta_y \rho\rangle}\approx\frac{gC D^2}{f}\left(\beta\delta\langle \Delta_y S\rangle -\alpha\delta\langle \Delta_y T\rangle \right)=\delta\{ \hat{\Psi}_{\mathrm{int}}\}_{\langle \Delta_y S\rangle}+\delta\{\hat{\Psi}_{\mathrm{int}}\}_{\langle \Delta_y T\rangle}
	\label{PD_Decomp}
\end{equation}
Here, $\langle \Delta_y T \rangle$ and $\langle \Delta_y S \rangle$ denote the vertically-averaged meridional temperature and salinity contrasts, respectively.
The coefficients $\alpha$ (= 1.3$\times$10$^{-4}$ K$^{-1}$) and $\beta$ (= 7.6$\times$10$^{-4}$ g$^{-1}$ kg) are determined by regressing $\delta\{\langle \Delta_y \rho \rangle\}_{\langle \Delta_y T \rangle}$ and $\delta\{\langle \Delta_y \rho \rangle\}_{\langle \Delta_y S \rangle}$ onto $\delta \langle \Delta_y T \rangle$ and $\delta \langle \Delta_y S \rangle$, respectively. 
The errors introduced by using a linear equation of state are evaluated by computing $\delta\{ \hat{\Psi}_{\mathrm{int}}\}_{\langle \Delta_y \rho\rangle}-(\delta\{ \hat{\Psi}_{\mathrm{int}}\}_{\langle \Delta_y S\rangle}+\delta\{\hat{\Psi}_{\mathrm{int}}\}_{\langle \Delta_y T\rangle})$, which were verified to be negligible ($\leq$ 0.1 Sv).

\subsection{Freshwater and Heat balance}

The drivers of temperature and salinity variations  affecting  the AMOC strength are analyzed using a heat and freshwater 
budget analysis. For an arbitrary Atlantic sector, $\mathcal{A}$ (bounded by latitudes $y_1$ and $y_2$), the freshwater 
and heat balance equations are \citep{drijfhout2011stability}:
\begin{subequations}
	\label{Cons_Eq}
	\renewcommand{\theequation}{\theparentequation\alph{equation}}
	\begin{align}
		\frac{d W}{dt} &=  \Delta F_{\mathrm{\mathrm{ov}}} + \Delta F_{\mathrm{az}} + F_{\mathrm{res}} +  F_{\mathrm{surf}} \label{W_Eq}\\
		\frac{d \mathrm{OHC}}{dt} &=  \Delta Q_{\mathrm{\mathrm{ov}}} + \Delta Q_{\mathrm{az}} + Q_{\mathrm{res}}+  Q_{\mathrm{surf}} \label{Q_Eq}\
	\end{align}
	\label{Balance}
\end{subequations}
Here, $W$ and OHC represent the volume-integrated freshwater and ocean heat content over the Atlantic sector $\mathcal{A}$. 
The operator $\Delta$ denotes the difference in horizontal freshwater and heat transport between the southern ($y_1$) and northern ($y_2$) boundaries of $\mathcal{A}$.

The terms $F_{\mathrm{\mathrm{ov}}}$ ($Q_{\mathrm{\mathrm{ov}}}$) and $F_{\mathrm{\mathrm{az}}}$ ($Q_{\mathrm{\mathrm{az}}}$) represent the freshwater (heat) transport by the overturning and gyre circulation, respectively. They are defined as:
\begin{subequations}
	\begin{align}
		F_{\mathrm{ov}}(y)  = - \frac{1}{S_0}\int_{-H}^0\left[\int_{x_W}^{x_E}v^* dx\right](\hat {S} -S_0)dz, &&	F_{\mathrm{az}}(y)  = - \frac{1}{S_0} \int_{-H}^0\left[\int_{x_W}^{x_E} v'S'dx\right]dz.
		\label{FovFazEq}\\
		Q_{\mathrm{ov}}(y)  = c_p \rho_0 \int_{-H}^0\left[\int_{x_W}^{x_E}v^* dx\right] \hat {T} dz, &&	Q_{\mathrm{az}}(y)  = c_p \rho_0 \int_{-H}^0\left[\int_{x_W}^{x_E} v'T'dx\right]dz.
		\label{QovFazEq}
	\end{align}
\end{subequations}
where $v^*$ represents the difference between the zonally-averaged (hatted quantities) and section-averaged velocity, primed quantities indicate deviations from the zonal averages.
The constant $c_p$ is the specific heat capacity of seawater (4186 J kg$^{-1}$ K$^{-1}$) and $S_0$ is a reference salinity set to 35 g kg$^{-1}$. 

The terms $F_{\mathrm{surf}}$ and $Q_{\mathrm{surf}}$ represent the area-integrated surface freshwater and heat fluxes for the Atlantic sector, $\mathcal{A}$. 
In the CESM, $F_{\mathrm{surf}}=P-E+R+I+M+F_H$ accounts for precipitation ($P$), evaporation ($E$, directed to the atmosphere), land run-off ($R$), ice ($I$), sea-ice melt and formation ($M$), and the imposed hosing surface freshwater flux ($F_H$). 
Similarly, $Q_{\mathrm{surf}}=SW+LW+SH+LH+MH$ includes short-wave radiation ($SW$), long-wave radiation ($LW$), latent heat flux ($LH$), sensible heat flux ($SH$), and heat flux due to sea-ice melting and freezing ($MH$). 
When $Q_{\mathrm{surf}}<0$, the heat flux is directed from the ocean to the atmosphere. 
Any contributions from mixing, barotropic transport, and diffusion are considered as the residual freshwater flux ($F_{\mathrm{res}}$) and residual heat flux ($Q_{\mathrm{res}}$).

\section{Result}\label{Results}
\subsection{Reconstructing the AMOC}\label{Recon}

\begin{figure*}[htpb]
	\captionsetup{justification=centering}
	\centering
	{\includegraphics[width=\textwidth]{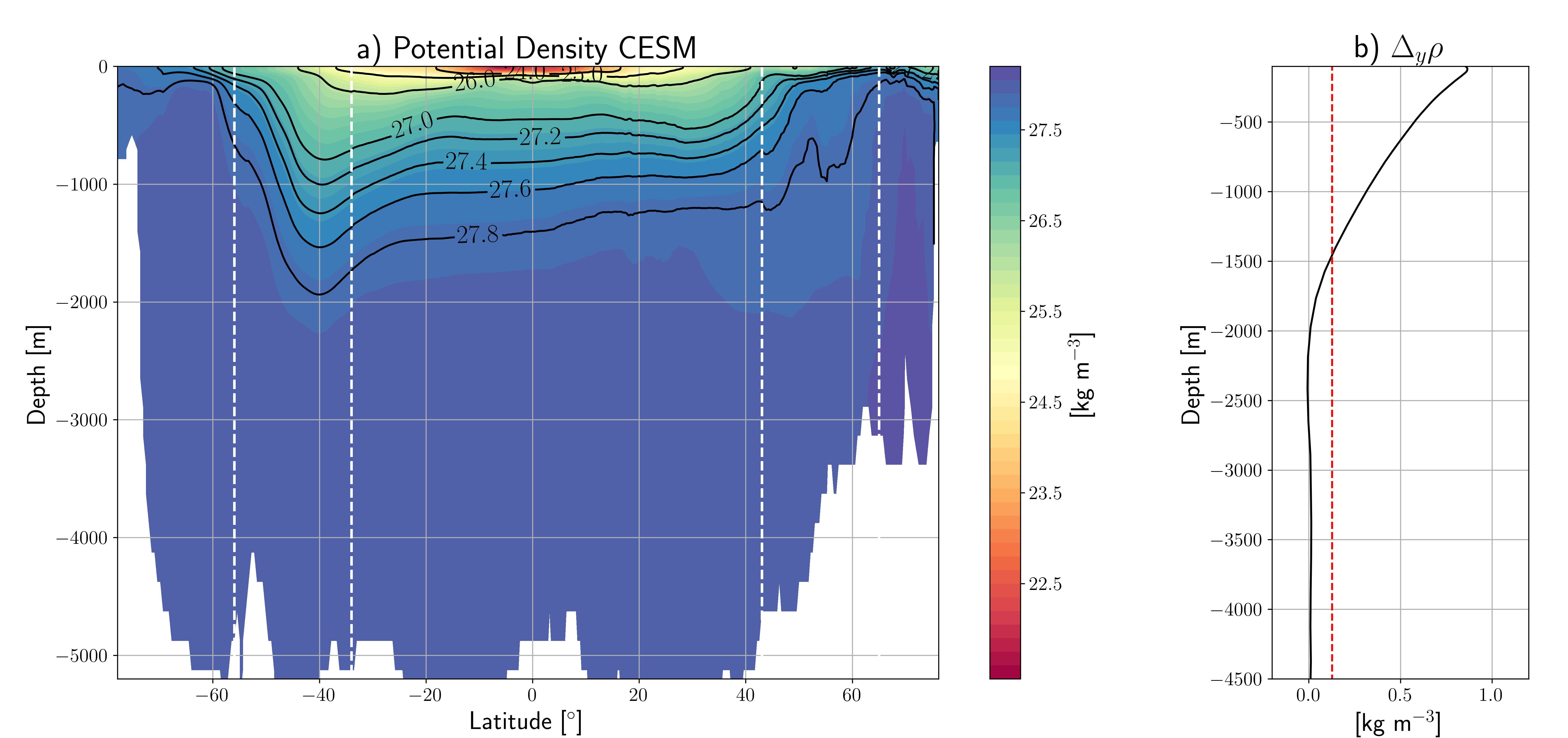}}
	\caption{\em \small (a): Time-mean and zonally-averaged potential density of the  Atlantic Ocean (full zonal extent) and Southern Ocean (zonal extent between 53$^\circ$W -- 20$^\circ$E), for the first 100 model years. (b): Time-mean meridional potential density with depth ($\Delta_y \rho$ thick black) and vertical average ($\langle\Delta_y \rho\rangle$ dashed red).}
	\label{F:Fig1a}
	\label{F:Fig1b}
	\label{F:Fig1}
\end{figure*}

To reconstruct the interior AMOC streamfunction, we require a precise definition of $\Delta_y \rho$ (Equation (\ref{TWB})).
In our study, $\Delta_y \rho(z) = \rho_n(z) - \rho_s(z)$ is defined as the meridional density difference between the North Atlantic ($\rho_n$) and the South Atlantic ($\rho_s$), with the latter potentially including parts of the Southern Ocean. 
The region in the North Atlantic that is crucial in determining AMOC strength lies within the latitude range where isopycnals, shared with the Antarctic Circumpolar Current (ACC), outcrop, and water mass transformation occurs \citep{nikurashin2012theory, wolfe2011adiabatic}. 
In the CESM, this region is bounded by the latitude range 43$^\circ$N –- 65$^\circ$N (Figure \ref{F:Fig1a}), with $\rho_n(z)$ calculated as the area-averaged potential density within this band. 
Our results remain robust with slight variations in the latitude range of $\rho_n$, which aligns with previous studies \citep{bonan2022transient, jansen2018transient}.

There is no consensus on the optimal latitude range for $\rho_s$ \citep{haskins2019explaining, haskins2020temperature, bonan2022transient, butler2016reconstructing} and here we aim to optimize the quality of the AMOC reconstruction by selecting a latitude range for $\rho_s$ that effectively captures the relative change in AMOC strength through the thermal wind balance. 
This optimal reconstruction is achieved when $\rho_s(z)$ is computed as the area-averaged potential density from 56$^\circ$S to 34$^\circ$S, and zonally between 53$^\circ$W and 20$^\circ$E.
The significance of this region is supported by physical arguments presented in \cite{wolfe2010sets} and \cite{nikurashin2011theory}, showing that the Atlantic interior stratification, in the adiabatic limit, is predominately influenced by processes in the Southern Ocean.
Furthermore, in the context of AMOC stability, \cite{wolfe2015multiple} and \cite{cimatoribus2014meridional} highlight the relevance of outcropping isopycnals in the Southern Ocean.  The chosen latitude range for $\rho_s$ captures the outcropping of overlapping 
isopycnals in the ACC (Figure~\ref{F:Fig1a}a). 

Figure~\ref{F:Fig1b}b shows the time-mean density contrast ($\Delta_y \rho(z)$) in the first 100 model years. 
Over the upper 2,000 m, the water mass is denser ($\Delta_y \rho(z)>0$) in the North Atlantic compared to the South Atlantic.
This positive density difference, drives a northward flow in the upper layers (equation (\ref{TWB})). 
There is hardly any density difference below 2,000~m depths.

\begin{figure*}[htpb]
	\captionsetup{justification=centering}
	\centering
	\includegraphics[width=\textwidth]{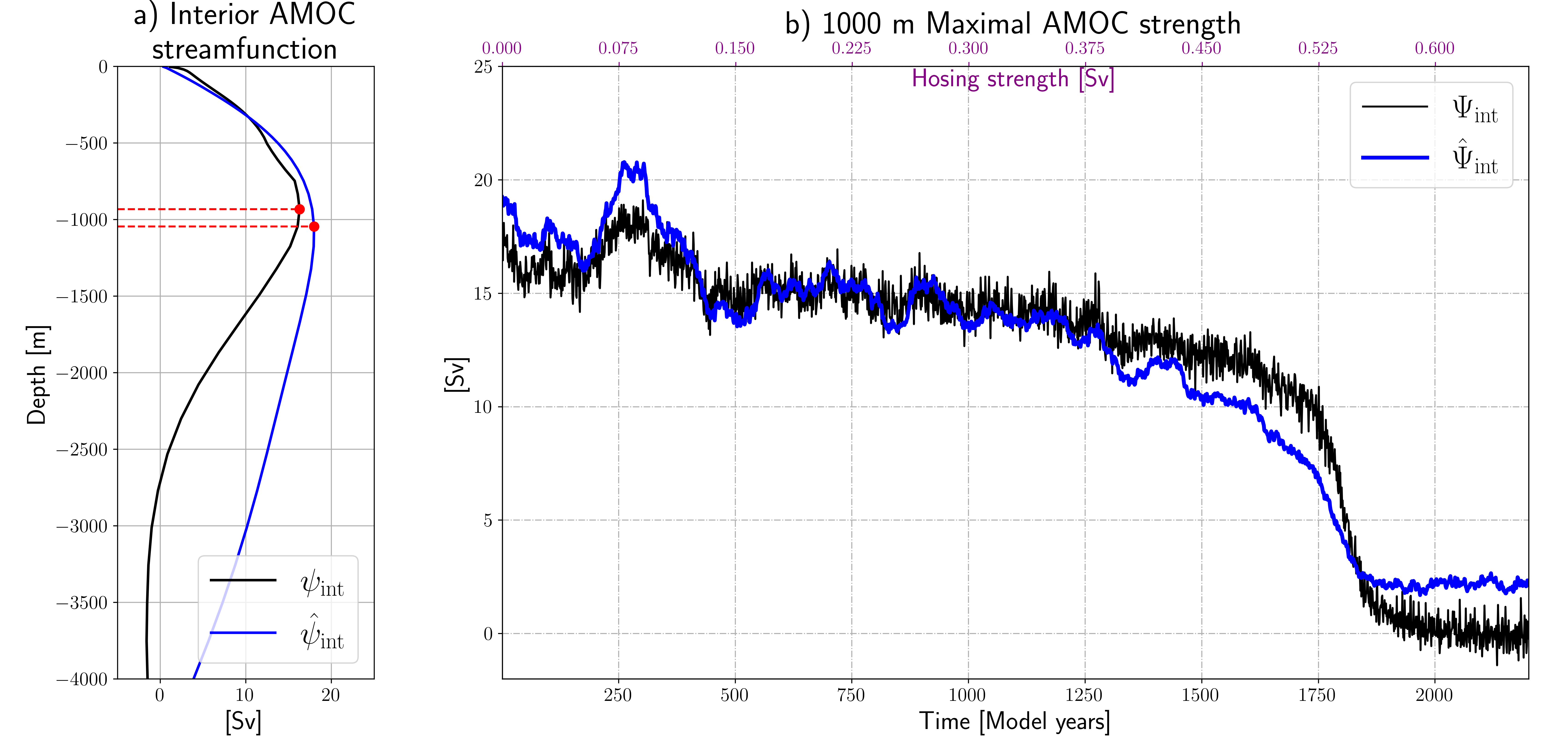}
	\caption{\em \small (a): Time-mean depth dependence of the interior overturning streamfunction (black curve) and TWB reconstruction (blue curve) over the first 100 model years. (b): Interior AMOC strength at 1,000 m under varying hosing strengths $F_H$, as simulated by the CESM (thick black curve) and the TWB reconstruction (thick blue curve).}
	\label{F:Fig2a}
	\label{F:Fig2b}
	\label{F:Fig2}
\end{figure*}

The dimensionless proportionality constant $C$ is determined by regressing the model-derived interior 
streamfunction ($\Psi_{\mathrm{int}}$) onto the TWB reconstruction ($\hat{\Psi}_{\mathrm{int}}$) yielding $C = 0.6$. Figure \ref{F:Fig2a}a shows the time-mean profile of the buoyancy reconstructed AMOC strength 
$\hat{\psi}_{\mathrm{int}}$  for the first 100 model years and the value of $\psi_{\mathrm{int}}$ directly 
determined from the CESM.  The TWB reconstruction reasonably 
captures the upper 1,000 m structure 
of $\psi_{\mathrm{int}}$. The $\psi_{\mathrm{int}}$ maximum is found at 950 m and the TWB reconstructed 
maxima is found 80 m lower. Below 1,000 m, $\psi_{\mathrm{int}}$ rapidly declines towards a  -1 Sv near 
the bottom. This residual corresponds to the Bering Strait volume transport \citep{juling2020atlantic}. The 
rapid decline is underestimated by $\hat{\psi}_{\mathrm{int}}$,  a discrepancy that arises due to the relatively 
large density differences between $\rho_n$ and $\rho_s$ below 1,000~m.  Additionally, $\hat{\psi}_{\mathrm{int}}$ 
is zero at the bottom (boundary conditions) and does not account for the volume transport through the 
Bering Strait.  

Figure \ref{F:Fig2b}b shows the AMOC strength $\hat{\Psi}_{\mathrm{int}}(t)$ and 
$\Psi_{\mathrm{int}}(t)$  over the complete 
simulation.  As reported by \cite{van2024physics}, the AMOC strength diminishes with increasing freshwater 
hosing, with the AMOC tipping event around $F_H$ = 0.53~Sv (around model year~1758). 
The AMOC displays significant variability in the first 500 years of hosing. \cite{van2024physics} attributed 
this to natural variability, but it may also be influenced by the imposed forcing  \citep{ehlert2014mechanism}.
The TWB reconstruction accurately captures the AMOC response to the hosing forcing, including multi-centennial 
variability, the extent of AMOC weakening due to hosing (approximately 17 Sv), the AMOC tipping event, and 
the stabilization of the collapsed AMOC.  This  accurate TWB reconstruction of the collapsed AMOC is achieved 
only when $\rho_s$ is computed south of the Atlantic basin boundary (34$^\circ$S), underlining the crucial role 
of Southern Ocean density stratification in AMOC stability \citep{wolfe2015multiple}.  However, some inaccuracies 
in the TWB reconstruction are notable: it slightly overestimates the rate of AMOC weakening and indicates that 
the AMOC off-state maintains a northward volume transport of 2.5 Sv.
%

\subsection{Attributing changes in AMOC strength}\label{ACA}

\begin{figure*}[htpb]
	\captionsetup{justification=centering}
	\centering
	\includegraphics[width=.5\textwidth]{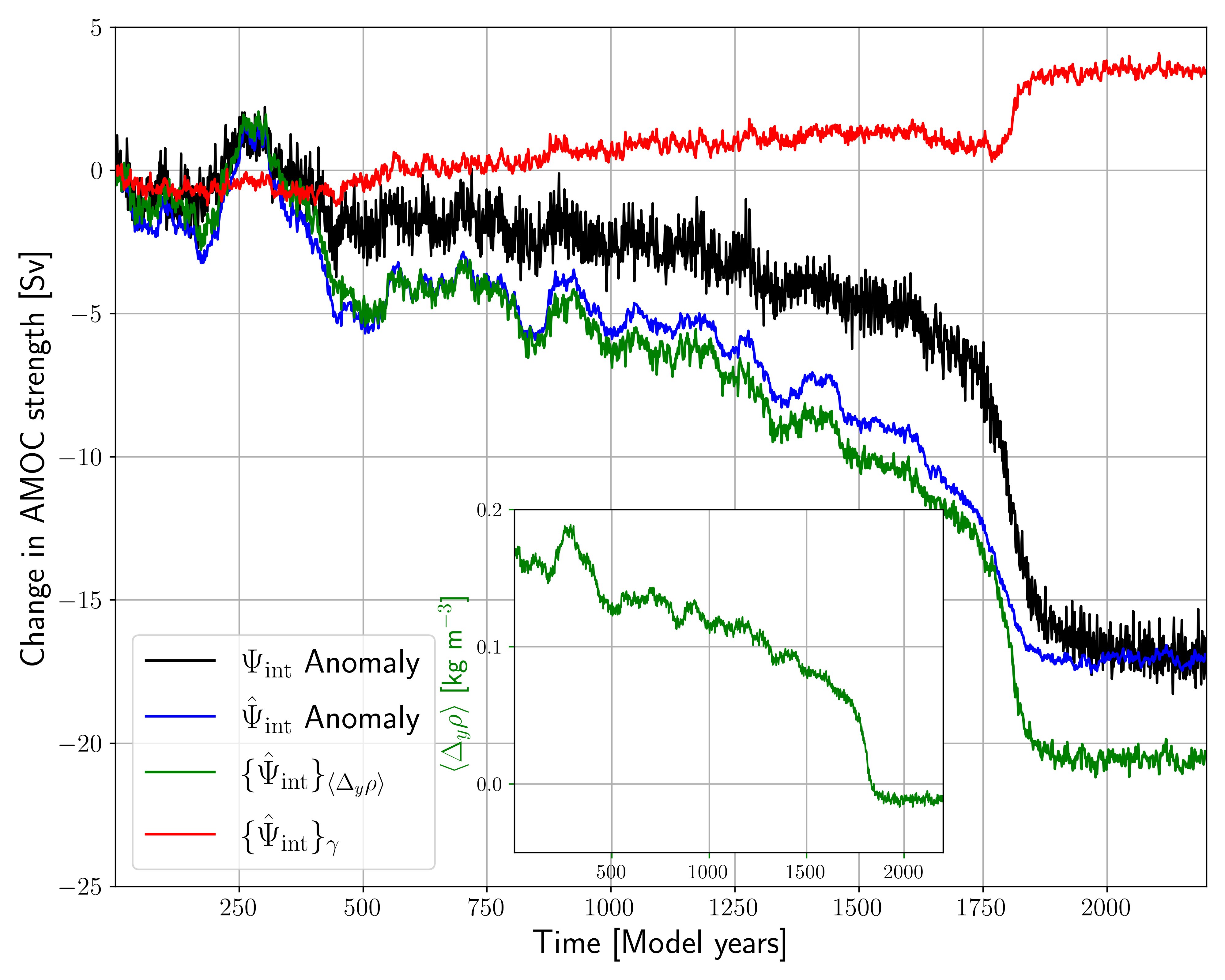}
	\caption{\em \small Decomposition of change in $\hat{\Psi}_{\mathrm{int}}$ (compared to the initial state) driven 
		by changes in vertically-averaged 
		density contrast (thick green line) and stratification (thick red line). Inset shows time series of 
		$\langle\Delta_y\rho\rangle$ (thick green line).}
	\label{F:Fig3}
\end{figure*}

Given that the TWB reconstructed AMOC strength $\hat{\Psi}_{\mathrm{int}}$ provides a reliable estimate for $\Psi_{\mathrm{int}}$, 
we can interpret changes in AMOC strength as primarily resulting from changes in the basin-scale meridional 
potential density contrast, $\Delta_y \rho(z)$.  Applying Equation (\ref{PerturbationPsi}), we decompose AMOC 
changes into components related to the vertically-averaged  potential density field  (i.e. $\delta \{\Psi_{\mathrm{int}}\}_{\langle \Delta_y \rho \rangle}$)   and to variations in stratification  (i.e., $\delta \{\Psi_{\mathrm{int}}\}_{\gamma}$). 
The vertically-averaged density contrast is responsible for an AMOC weakening of approximately 
20 Sv (Figure \ref{F:Fig3}).   Changes in $\gamma$ contribute to a minor AMOC strengthening of 
about 3 Sv, which can be attributed to the deepening of the pycnocline.  
%
Changes in the vertically-averaged potential density contrast (see inset Figure \ref{F:Fig3}) are therefore dominating the AMOC decline, and the contribution by $\delta\{\Psi_{\mathrm{int}}\}_{\gamma}$ is opposite and much smaller. 
Notably, $\langle\Delta_y\rho\rangle$ approaches zero when the AMOC collapses which  is consistent with results from box models \citep{stommel1961thermohaline, rahmstorf1996freshwater, cimatoribus2014meridional}. 

\begin{figure*}[htpb]
	\captionsetup{justification=centering}
	\centering
	\includegraphics[width=\textwidth]{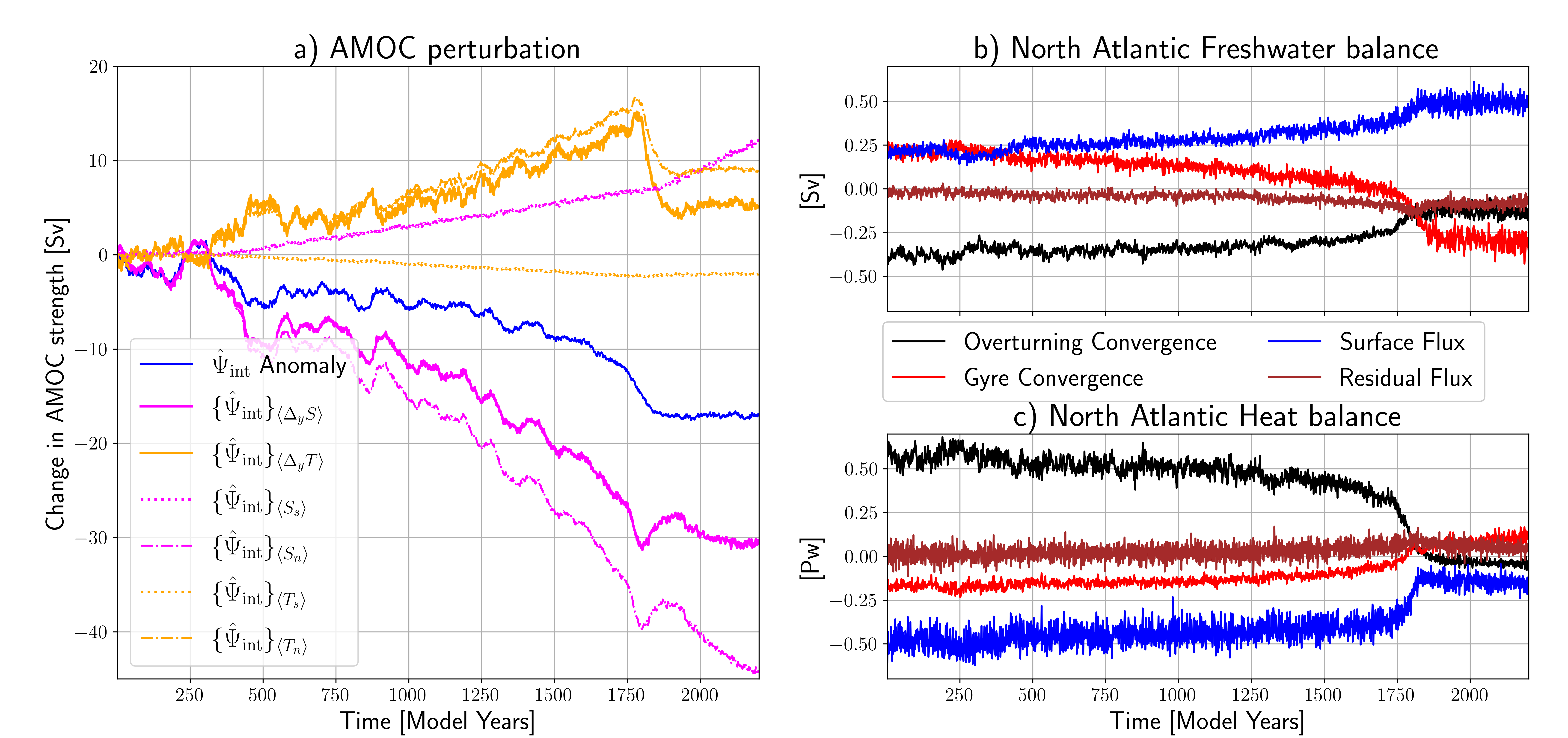}
	\caption{\em \small (a): Decomposition of AMOC perturbations by vertically-averaged potential density changes into thermal (orange) and haline (magenta) by the North (dash-dot) and South Atlantic (dotted).
		(b \& c): The North Atlantic freshwater and heat balances (see equation~(\ref{Balance})).}
	\label{F:Fig4a}
	\label{F:Fig4b}
	\label{F:Fig4c}
	\label{F:Fig4}
\end{figure*}

We further decompose AMOC changes by a salinity ($\langle\Delta_y S\rangle = \langle S_n - S_s\rangle$) and 
temperature ($\langle\Delta_y T\rangle = \langle T_n - T_s\rangle$) contribution in 
Figure \ref{F:Fig4a}a, showing  $\{\hat{\Psi}_{\mathrm{int}}\}_{\langle\Delta_y S\rangle}$ and 
$\{\hat{\Psi}_{\mathrm{int}}\}_{\langle\Delta_y T\rangle}$, respectively.  
Under the increasing freshwater flux forcing, a declined vertically-averaged salinity contrast is weakening 
the AMOC ($\{\hat{\Psi}_{\mathrm{int}}\}_{\langle\Delta_y S\rangle}$), 
while temperature variations ( $\{\hat{\Psi}_{\mathrm{int}}\}_{\langle\Delta_y T\rangle}$) cause  an opposite AMOC response which is approximately half the magnitude of the salinity-driven decrease. 
Between model years 1,800 and 1,900 the contributions reverse and the temperature contrast drives an AMOC decline of 6~Sv while the salinity contrast offsets this decline by about 2~Sv. After model year 2,000, both the salinity and temperature fields stabilize.

Further decomposing the AMOC response into North Atlantic and South Atlantic contributions (Figure \ref{F:Fig4}a), we find that vertically-averaged freshening in the North Atlantic ($\{\hat{\Psi}_{\mathrm{int}}\}_{S_n}$) 
is the primary driver of the decreasing AMOC strength prior to model year 1,800. 
Figure \ref{F:Fig4}b shows that this freshening results from an increased surface freshwater flux and freshwater convergence by the overturning circulation. 
Vertically-averaged cooling of the North Atlantic basin ($\{\hat{\Psi}_{\mathrm{int}}\}_{T_n}$) provides a 50$\%$ offset to this decreasing tendency and is related to a decreased overturning heat convergence (Figure \ref{F:Fig4}c), 
leading to the characteristic AMOC fingerprint \citep{drijfhout2012decline,caesar2018observed}.

The AMOC responses induced by the South Atlantic partly offsets the AMOC weakening by the North Atlantic. 
The main contribution in the South Atlantic comes from vertically-averaged freshening ($\{\hat{\Psi}_{\mathrm{int}}\}_{\langle S_s \rangle}$),
the vertically-averaged cooling ($\{\hat{\Psi}_{\mathrm{int}}\}_{\langle T_s \rangle}$) slightly reduces the AMOC strength. 
The combined South Atlantic response ($\{\hat{\Psi}_{\mathrm{int}}\}_{\langle \rho_s \rangle}$) is +20\% (up to the tipping event) compared to the combined North Atlantic response ($\{\hat{\Psi}_{\mathrm{int}}\}_{\langle \rho_n \rangle}$), 
illustrating that North Atlantic salinity changes  primarily drive the AMOC decline.

Both the vertically-averaged temperature and salinity over the North Atlantic region increase after model year~1,800.
The surface heat flux and heat transport by the gyre drive this ocean heat convergence (Figure \ref{F:Fig4c}c),
where the former is driven by latent, sensible, and longwave heat fluxes changes in response to an enhanced North Atlantic sea-ice cover \citep{van2023asymmetry}.
This temperature increase results in AMOC weakening after model year~1,800 and stabilizes the collapsed AMOC state. 
The increase in North Atlantic salinity is related to a decreased gyre freshwater convergence and
eventually equilibrates with the overturning and surface freshwater fluxes by model year~2,100.

\subsection{AMOC feedbacks: A box-model perspective}\label{QAF}

We have demonstrated that a realistic AMOC reconstruction can be obtained by using only vertically-averaged quantities.
This is analogous to box models where tracer quantities are assumed to be well mixed within each box 
\citep{rahmstorf1996freshwater}.  Therefore, it is tempting to capture the AMOC responses in the CESM with  a box model 
approach and  analyse the processes (i.e., salinity) driving the AMOC weakening under the increasing freshwater flux forcing.
This will be done by analysing the steady-state Atlantic freshwater budget between 34$^{\circ}$S and 
65$^{\circ}$N (Equation (\ref{W_Eq})): 
\begin{equation}	0=F_{\mathrm{surf}}+F_{\mathrm{res}}+\left(F_{\mathrm{ovS}}-F_{\mathrm{ovN}}\right)+
	\left(F_{\mathrm{azS}}-F_{\mathrm{azN}}\right). 
	\label{Steady_State_Balance}
\end{equation}
Here, the quantities with a subscript~S (N) are evaluated at 34$^{\circ}$S (65$^{\circ}$N) and the other components are integrated 
over the Atlantic Ocean surface between 34$^{\circ}$S to 65$^{\circ}$N.

\begin{figure*}[htpb]
	\captionsetup{justification=centering}
	\centering
	\includegraphics[width=\textwidth]{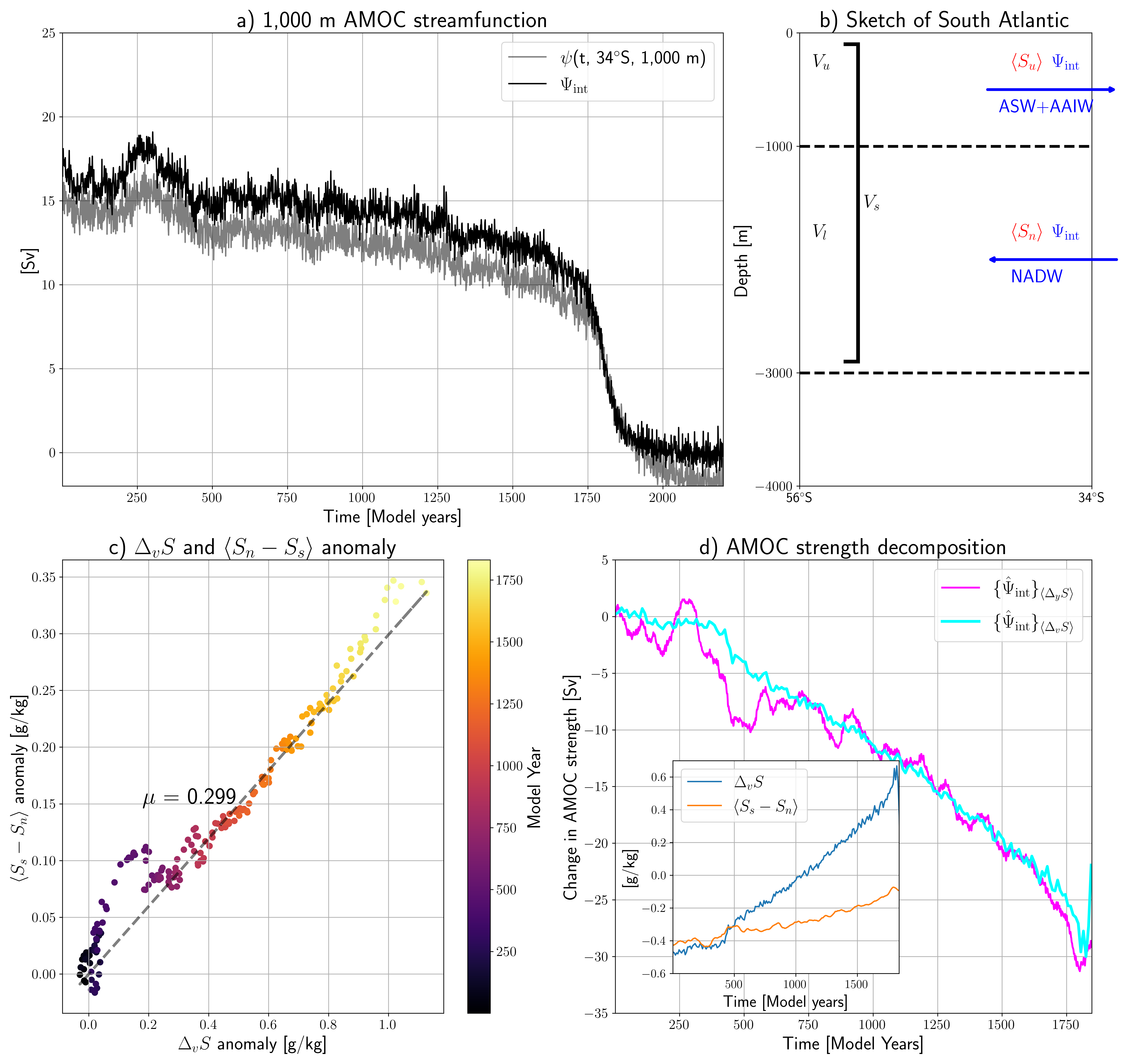}
	\caption{\em \small (a): Interior AMOC streamfunction (thick black) and $\psi$ (equation (\ref{MStrF})) evaluated at 34$^\circ$S and 1,000 depth (thin black). (b): Sketch illustrating the rationale behind equation (\ref{Physical_Rel}). Arrows indicate the volume transport of the overturning circulation, with the abbreviation below each arrow corresponding to the transported Atlantic water mass. (c): Relation between $\Delta_v S$ and $\langle S_s-S_n\rangle$ anomaly. The slope $\mu$ is calculated from a linear regression. Each dot represents a 10-year average. (d): AMOC anomaly resulting from change in $\Delta_v S$ (thick cyan) and $\langle\Delta_y S\rangle$ (thick magenta). Inset shows that time series of $\Delta_v S$ and $\langle S_s-S_n\rangle$.}
	\label{F:Fig5a}
	\label{F:Fig5b}
	\label{F:Fig5c}
	\label{F:Fig5d}
	\label{F:Fig5}
\end{figure*}

Given that the AMOC strength at 1,000 m depth at 34$^\circ$S closely resembles $\Psi_{\mathrm{int}}$ (Figure \ref{F:Fig5a}a) and considering the negligible contribution of Antarctic Bottom Water (AABW) to $F_{\mathrm{ovS}}$ \citep{van2024persistent}, we express $F_{\mathrm{ovS}}$ as follows (with $\Psi_{\mathrm{int}} > 0$):
\begin{equation} 
	F_{\mathrm{ovS}} = -\frac{\Psi_{\mathrm{int}} \Delta_{v}S}{S_0}. 
	\label{FovS_def}
\end{equation}
where $\Delta_{v}S$ (determined below)  and denotes the velocity-weighted salinity contrast between the 
northward and southward flowing limbs of the AMOC at 34°S \citep{sijp2012characterising}.  The $F_{\mathrm{ovS}}$ 
sign is now directly related to the $\Delta_{v}S$ sign. It is important to note that Equation (\ref{FovS_def}) is only 
valid when a northward overturning state exists, and thus our analysis applies this equation only up to model 
year 1,850. 

In simple box models, the velocity-weighted salinity contrast is simply $\Delta_{v}S=S_s-S_n$ and assumes that $\delta S_s$ is uniform throughout the water column \citep{rahmstorf1996freshwater}.
This is not the case for the CESM. 
Specifically, freshwater anomalies originating from the North Atlantic accumulate in the North Atlantic Deep Water (NADW) and are transported southward (quasi-adiabatically) to the South Atlantic (Figure \ref{F:Fig5}b). 
Hence, it is reasonable to expect that the $\delta \langle S_n\rangle\approx\delta \langle S_l\rangle$, where $S_l$ represents the  volume averaged salinity of the South Atlantic NADW. The equilibrated change in South Atlantic salinity can therefore be expressed as:
\begin{equation}
	\delta \langle S_s\rangle  \approx \frac{\delta \langle S_n\rangle V_l + \delta S_u V_u}{V_s}
	\label{Physical_Rel}
\end{equation}
where $V_s = V_l + V_u$, with $V_l$ representing the volume of the lower limb in the South Atlantic and $V_u$ representing the volume of the upper limb. Here, $S_u$ is the volume-averaged salinity over $V_u$, representing the Antarctic Intermediate Water (AAIW) and Atlantic Surface Water (ASW). 
Considering that $\frac{V_u}{V_s} \approx 0.25 - 0.3$ (Figure~\ref{F:Fig5}b), it follows that $\delta \Delta_v S = \delta (S_u - \langle S_n \rangle)$. 
Therefore, we express $\delta \Delta_y \langle S \rangle$ as $-\mu \delta \Delta_v S$, with $\mu \approx 0.25 - 0.3$. We find $\mu = 0.3$ for the CESM (Figure~\ref{F:Fig5a}c), 
demonstrating that this approximation and quasi-equilibrium hosing conditions hold. 
This implies that AMOC responses can be directly related to the vertically-averaged salinity contrast, 
where we now use $\mu\delta\Delta_v S$ instead of $\delta\Delta_y \langle S\rangle$ (Figure~\ref{F:Fig5b}d). The $\mu\delta\Delta_v S$ contribution captures the overall AMOC decline prior to the tipping event.

Applying a small perturbation to the Atlantic freshwater balance (Equation (\ref{Steady_State_Balance})) around a reference
state and using Equation (\ref{FovS_def}), we can rewrite it to express $\delta \Delta_v S$ as follows:
\begin{equation}
	\delta \Delta_v S = \frac{S_0}{\Psi_\mathrm{int}}\left(\delta\left(F_{\mathrm{azS}}-F_{\mathrm{azN}}\right)-\delta F_{\mathrm{ovN}}-\frac{1}{S_0}\Delta_v S\delta \Psi_{\mathrm{int}}+\delta F_{\mathrm{surf}}+\delta F_{\mathrm{res}}\right).
	\label{Perturbation_DeltavS}
\end{equation}
By combining Equation (\ref{PD_Decomp}) and (\ref{Perturbation_DeltavS}) we obtain:
\begin{equation}
	\begin{split}
		\delta \{\hat{\Psi}_{\mathrm{int}}\}_{\Delta_v S} \approx  \frac{-g\mu S_0 C\beta D^2}{\Psi_\mathrm{int}f}\left(\delta\left(F_{\mathrm{azS}}-F_{\mathrm{azN}}\right)-\delta F_{\mathrm{ovN}}-\frac{1}{S_0}\delta \Psi_{\mathrm{int}}\Delta_v S+\delta F_{\mathrm{surf}}+\delta F_{\mathrm{res}}\right),\\
		\approx \delta \{\hat{\Psi}_{\mathrm{int}}\}_{\mathrm{Gyre}}+\delta \{\hat{\Psi}_{\mathrm{int}}\}_{\mathrm{\mathrm{ovN}}}+\delta \{\hat{\Psi}_{\mathrm{int}}\}_{\mathrm{Saf}}+\delta\{\hat{\Psi}_{\mathrm{int}}\}_{\mathrm{Surf}}+
		\delta\{\hat{\Psi}_{\mathrm{int}}\}_{\mathrm{Res}}
	\end{split}
	\label{Process_Decomp}
\end{equation}
Terms on the right hand side of Equation (\ref{Process_Decomp}) quantify the AMOC change resulting from alterations in processes that drive changes in $\Delta_v S$. 
Specifically, changes in: the gyre freshwater circulation ($ \delta \{\hat{\Psi}_{\mathrm{int}}\}_{\mathrm{Gyre}}$),
the overturning freshwater transport evaluated at the northern boundary of the Atlantic ($\delta \{\hat{\Psi}_{\mathrm{int}}\}_{\mathrm{ovN}}$), 
the overturning salt-advection feedback ($\delta \{\hat{\Psi}_{\mathrm{int}}\}_{\mathrm{Saf}}$), 
the area-integrated surface freshwater flux ($\delta \{\hat{\Psi}_{\mathrm{int}}\}_{\mathrm{Surf}}$), 
and the residual freshwater flux ($\delta \{\hat{\Psi}_{\mathrm{int}}\}_{\mathrm{Res}}$). 
It is important to note that contributions from the Atlantic freshwater tendency are neglected in 
equation (\ref{Steady_State_Balance}), leading to a small offset between the left and right 
hand side of the Equation.
Equation (\ref{Process_Decomp}) demonstrates that AMOC changes, and consequently AMOC stability, 
are constrained by alterations in the terms of the Atlantic-integrated freshwater balance, as originally proposed by \cite{rahmstorf1996freshwater} and \cite{sijp2012characterising}.  
Additionally, it highlights that the significance of the overturning salt-advection feedback, amid the presence of other feedbacks, depends on the changes in other components of the freshwater budget \citep{huisman2010indicator,gent2018commentary}. 

\begin{figure*}[htpb]
	\captionsetup{justification=centering}
	\centering
	\includegraphics[width=1\textwidth]{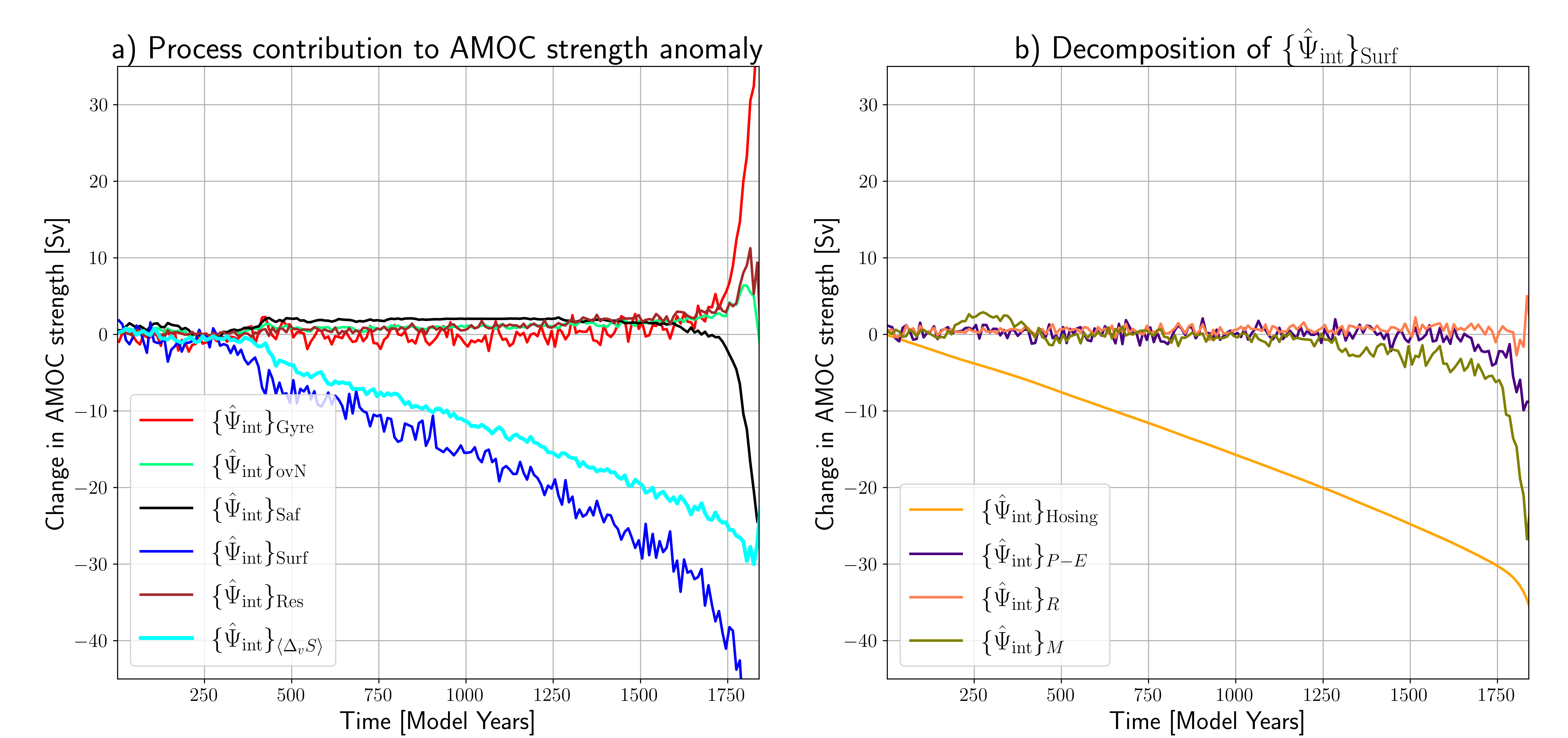}
	\caption{\em \small (a): Contribution of the different terms in equation (\ref{Process_Decomp}) to the total change in $\{\hat{\Psi}_{\mathrm{int}}\}_{\Delta_v S}$ (cyan line).
		(b): Decomposition of $\{\hat{\Psi}_{\mathrm{int}}\}_{\mathrm{Surf}}$ into its various components (Melt, $P-E$, hosing and land run-off). Contributions from ice-runoff are not shown, as they was found to be negligible.}
	\label{F:Fig6a}
	\label{F:Fig6b}
	\label{F:Fig6}
\end{figure*}

The different contributions to AMOC changes are shown in Figure~\ref{F:Fig6a}a and the dominant driver of AMOC weakening is the surface freshwater flux ($\{\hat{\Psi}_{\mathrm{int}}\}_{\mathrm{Surf}}$). 
As expected, the surface freshwater flux is primarily forced under the increasing hosing strength (Figure~\ref{F:Fig6b}b). 
The contribution by hosing is linearly declining up to the AMOC tipping event, but the $\{\hat{\Psi}_{\mathrm{int}}\}_{\mathrm{Surf}}$ decline is non linear after model year~1,200. 
This non-linear decline is related to additional ocean-atmosphere-sea ice feedbacks, which show a pronounced melt ($M$) and atmospheric surface freshwater flux ($P-E$) contribution. 
The contribution by river run-off ($R$) remains relatively small and ice run-off and brine rejection are negligible (not shown).

Changes in gyre circulation are minimal before model year 1,500, after which they counterbalance the AMOC decline near tipping (Figure \ref{F:Fig2b}b), contributing an increase of about 7 Sv from model year 1,000 to 1,750. 
Similar behavior is noted for $\delta \{\hat{\Psi}_{\mathrm{int}}\}_{\mathrm{Res}}$ and $\delta \{\hat{\Psi}_{\mathrm{int}}\}_{\mathrm{ovN}}$ which offset the AMOC decrease by roughly 3 Sv and 1.5 Sv, respectively, over the same period.
The overturning salt-advection feedback ($\{\hat{\Psi}_{\mathrm{int}}\}_{\mathrm{Saf}}$) initially stabilizes the AMOC strength by 2~Sv during the first 1,000 model years. 
Thereafter, the salt-advection feedback response contributes 6~Sv to AMOC weakening between model year~1,000 to 1,750.
The salt-advection feedback, sea-ice melt and ocean-atmosphere fluxes contribute to an accelerated AMOC decline,
where the salt-advection feedback is eventually the most dominant contribution. The precise mechanisms will be 
further investigated in the next section. 


\section{Analysis: Feedbacks in the CESM model}\label{Anl}
\subsection{The  Salt-Advection Feedback}

\begin{figure*}[htpb]
	\captionsetup{justification=centering}
	\centering
	\includegraphics[width=1\textwidth]{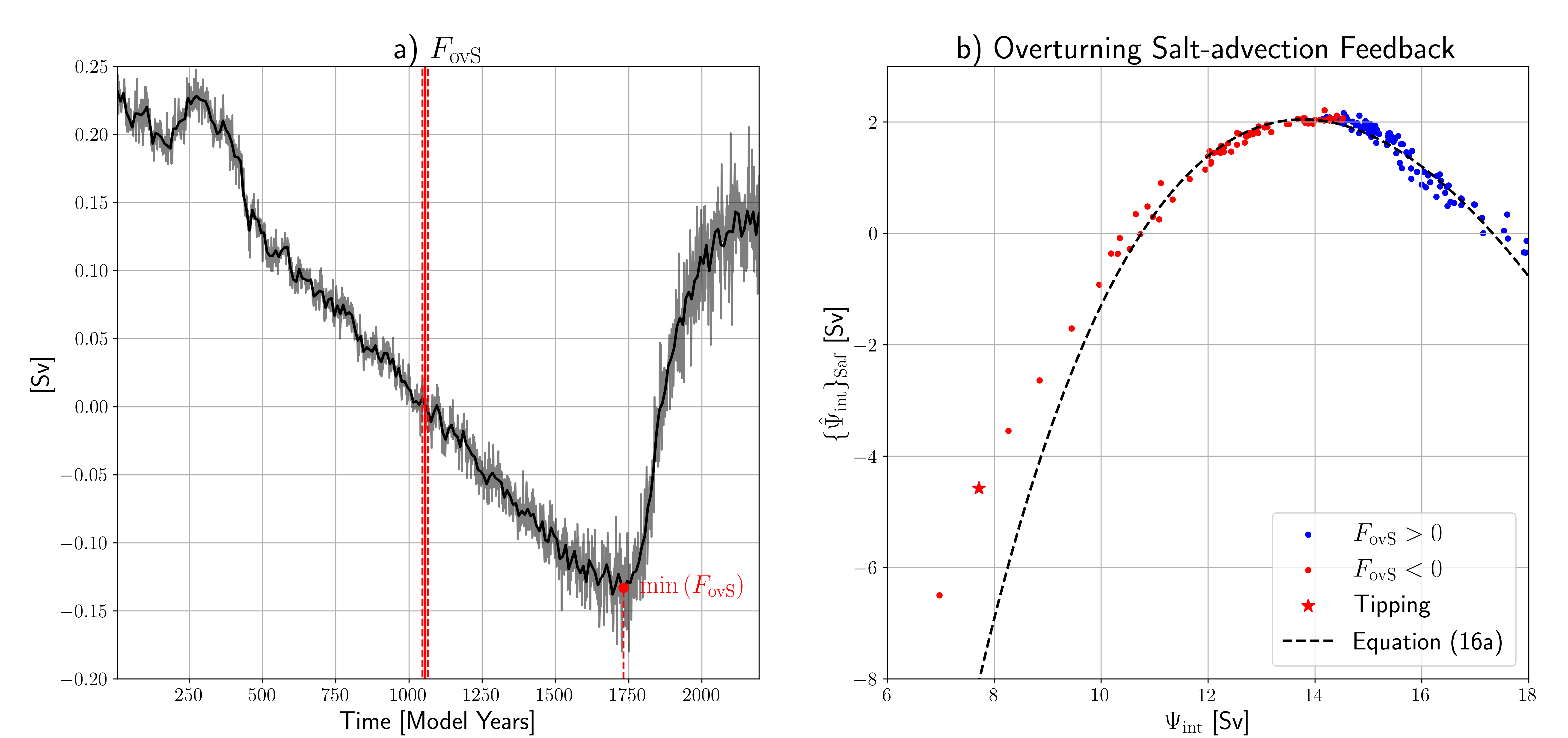}
	\caption{\em \small (a): Time series of $F_{\mathrm{ovS}}$ during the hosing simulation (thin black line), along with the 10-year moving average of the same time series (thick black line). The sign of $F_{\mathrm{ovS}}$ switches after 1,055 years of hosing (confidence interval: 1,046–1,064 years, 90\% confidence, indicated by red shaded bands). The minimum value of $F_{\mathrm{ovS}}$ is reached at year 1,732 (confidence interval: 1,727–1,740, 90\% confidence) \citep{van2024physics}.
	(b): The relation between $\{\hat{\Psi}_{\mathrm{int, r}}\}_{\mathrm{Saf}}$ and $\hat{\Psi}_{\mathrm{int}}$ for different $F_{\mathrm{ovS}}$ signs.}
	\label{F:Fig7a}
	\label{F:Fig7b}
	\label{F:Fig7}
\end{figure*}

The systematic decomposition (\ref{Process_Decomp})  showed that the overturning salt-advection feedback 
initially stabilises the AMOC during the first 1,000 model years and contributes to AMOC weakening and 
the collapse thereafter.  In Figure~\ref{F:Fig7b}a,  $F_{\mathrm{ovS}}$ is plotted for the complete CESM simulation 
showing that it changes sign (from positive to negative) around model year $1055$ and goes through 
a minimum just before the collapse. The relation between the salt-advection feedback response 
($\delta\{\hat{\Psi}_{\mathrm{int}}\}_{\mathrm{Saf}}$) and the AMOC strength is shown in 
Figure~\ref{F:Fig7b}b. When $F_{\mathrm{ovS}}>0$, a negative AMOC perturbation ($\delta\Psi_{\mathrm{int}}<0$) 
is  opposed by a positive salt-advection feedback response ($\delta\{\hat{\Psi}_{\mathrm{int}}\}_{\mathrm{Saf}} > 0$). 
However, when $F_{\mathrm{ovS}}<0$, a negative AMOC perturbation ($\delta\Psi_{\mathrm{int}}<0$) is further 
amplified by the salt-advection feedback response ($\delta\{\hat{\Psi}_{\mathrm{int}}\}_{\mathrm{Saf}} < 0$). 
The salt-advection feedback strength response becomes stronger (i.e., more negative) for a weaker AMOC.
Such a relation between AMOC and $F_{\mathrm{ovS}}$ (i.e., salt-advection feedback strength)
is also found  in the CMIP6 models for the pre-industrial (1850 -- 1899) and present-day (1994 -- 2020) period \citep{van2024persistent,vanwesten2024substantialrisk21stcentury}.

In Appendix A we argue that $\hat{\Psi}_{\mathrm{int}}$ can be approximated as a linear function of $\Delta_v S$, i.e. 
\begin{equation}
	\hat{\Psi}_{\mathrm{int}}(t)\approx \frac{gCD^2\beta\mu}{f}\left(a-(1-c_1)\left(\Delta_vS(t)\right)\right),
	\label{Psi_Approx}
\end{equation}
where the constant $c_1$ represents the efficiency of the overturning thermal advective feedback in mitigating 
the salinity-driven AMOC decline. As noted earlier, $c_1$ is approximately 0.5 before tipping occurs (Figure \ref{F:FigA1}). The 
parameter $a$ ($\sim$ 0.5) is defined in Appendix A and is determined by the equilibrium AMOC strength 
in the absence of hosing.  

Equation (\ref{Psi_Approx}) enables an explicit evaluation of the salt-advection 
feedback strength as a function of $\Delta_v S$ (see Appendix A). The resulting expression is: 
\begin{subequations}
	\begin{align}
		\{\hat{\Psi}_{\mathrm{int}}\}_{\mathrm{Saf}}(t) &\approx \frac{g\mu C \beta D^2}{f(1 - c_1)} \left( (1 - c_1)(\Delta_v S(t) - \Delta_v S(0)) + a \log\left(\frac{a-(1-c_1)\Delta_v S(t)}{a-(1-c_1)\Delta_v S(0)}\right) \right),\\
		\frac{\partial \{\hat{\Psi}_{\mathrm{int}}\}_{\mathrm{Saf}}}{\partial \Delta_v S} &\approx \frac{g\mu C \beta D^2}{f} \frac{(c_1-1)\Delta_v S}{(c_1-1)\Delta_v S+a},
	\end{align}
	\label{SAF_Sol} 
\end{subequations}
Equation (\ref{SAF_Sol}b) shows that for $\Delta_v S < 0$ ($F_{\mathrm{ovS}} > 0$), $\delta\{\hat{\Psi}_{\mathrm{int}}\}_{\mathrm{Saf}} > 0$. 
Conversely, for $\Delta_v S > 0$ ($F_{\mathrm{ovS}} < 0$), $\delta\{\hat{\Psi}_{\mathrm{int}}\}_{\mathrm{Saf}} < 0$. 
Moreover, for increasing values of $\Delta_v S$ (or $F_{\mathrm{ovS}}$), equation (\ref{SAF_Sol}b) reveals that the 
salt-advection feedback becomes logarithmically more effective at weakening the AMOC. 

Using equation (\ref{Psi_Approx}) to rewrite equation (\ref{SAF_Sol}a) in terms of $\hat{\Psi}_{\mathrm{int}}$, and substituting all necessary parameter values, we obtain an semi-analytic estimate for the salt-advection feedback strength as a function of $\hat{\Psi}_{\mathrm{int}}$ 
(the dashed curve in Figure \ref{F:Fig7}b). 
This estimate accurately captures the gradual increase in salt-advection feedback strength as $\Delta_v S$ rises. Despite some assumptions, the framework can recover the gradual AMOC destabilization resulting from the salt-advection feedback and its dependence on the $F_{\mathrm{ovS}}$.

\subsection{Atmospheric and sea-ice AMOC feedbacks}

Two additional contributions to the accelerated AMOC decline are the ocean-atmosphere and sea-ice melt surface freshwater fluxes (Figure \ref{F:Fig6b}b). Both fluxes increase during the hosing simulation, with sea-ice melt freshwater fluxes increasing by about 0.15~Sv from model year 1 to 1,800, representing a 150\% increase from the initial value, in particular after model year~1,200. Atmospheric freshwater fluxes increase by 0.09~Sv (7\% increase from initial value), with significant changes emerging after 1,600~model years. The sea-ice melt contribution is correlated with $\Psi_{\mathrm{int}}$, as confirmed by co-integration analysis \citep{engle1987co} with a confidence level of $p<0.01$. No significant co-integration was found between ocean-atmosphere freshwater fluxes and $\Psi_{\mathrm{int}}$.

\begin{figure*}[htpb]
	\captionsetup{justification=centering}
	\centering
	\includegraphics[width=.5\textwidth]{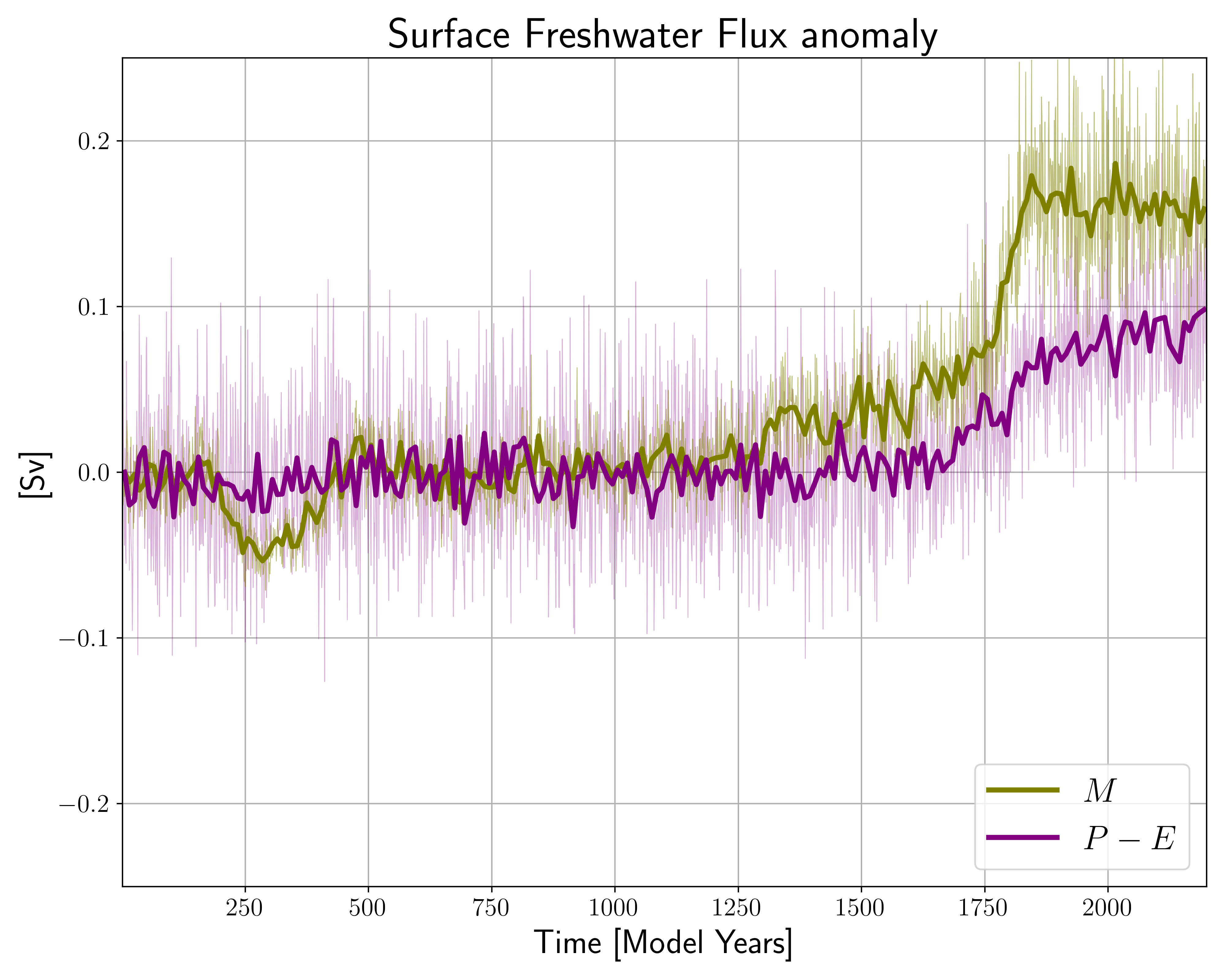}
	\caption{\em \small Time-series of anomalous atmospheric ($P-E$) and sea-ice melt/formation ($M$) surface freshwater fluxes over the Atlantic Ocean (34$^\circ$S -- 65$^\circ$N) in the hosing simulation. Thick lines represent 10-year average of thin lines.}
	\label{F:Fig8}
\end{figure*}

\begin{figure*}[htpb]
	\captionsetup{justification=centering}
	\centering
	\includegraphics[width=\textwidth]{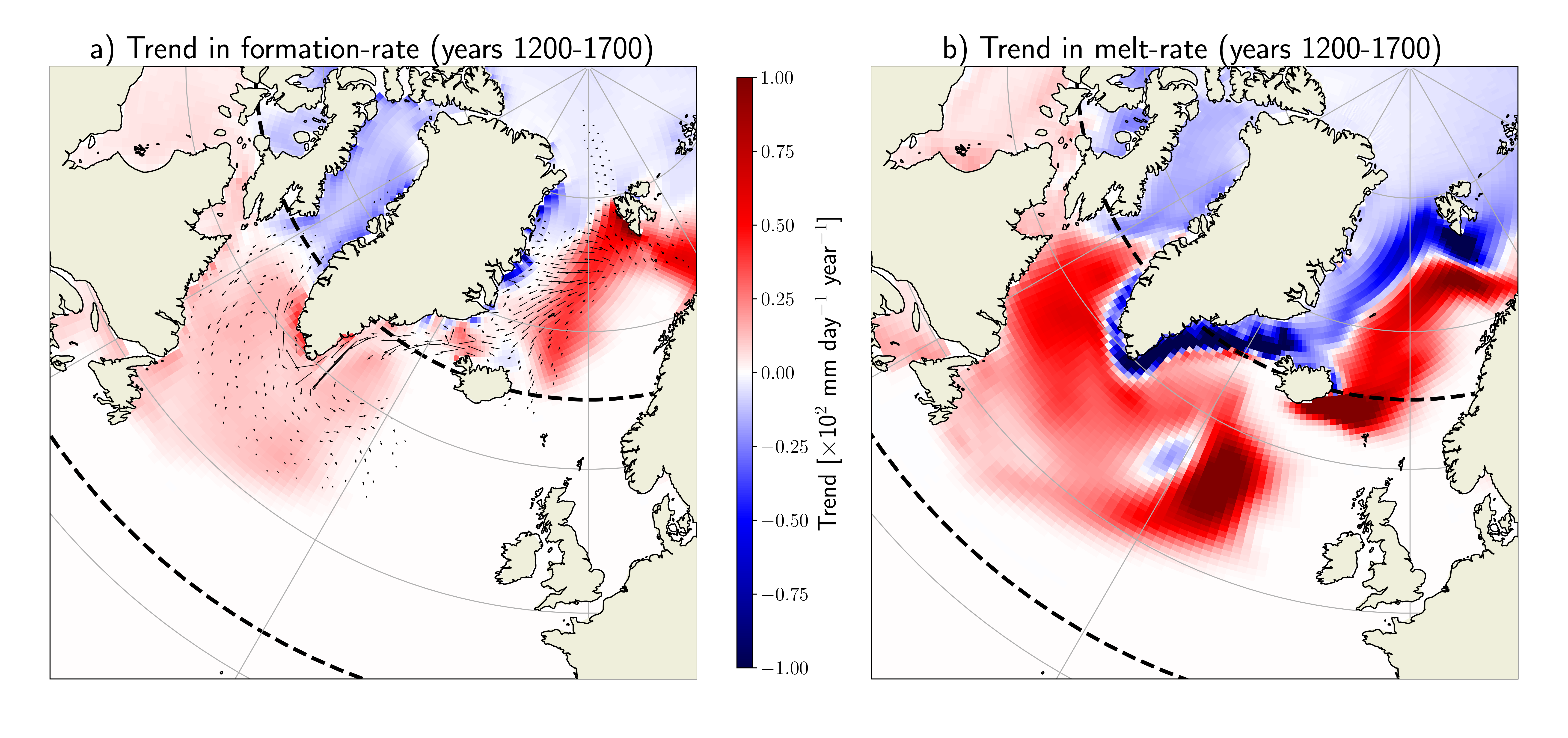}
	\caption{\em \small (a): Tendency of the yearly averaged sea-ice formation rate (colors) and ice transport (arrows) for years 1200 -- 1700. (b): Tendency of the yearly averaged melting rate for years 1200-1700. Only significant ($p<0.05$) trends are shown. Black dashed curves mark the boundaries of the North-Atlantic.}
	\label{F:Fig9a}
	\label{F:Fig9b}
	\label{F:Fig9}
\end{figure*}

In the CESM, the sea-ice melt freshwater flux can be either negative or positive. When $M$ is negative (during winter), sea-ice formation dominates over sea-ice melt, leading to freshwater being extracted from the ocean.  For positive $M$ (during spring and summer), sea-ice melt exceeds formation, resulting in an influx of freshwater into the ocean surface. The strong correlation between $\Psi_{\mathrm{int}}$ and $M$ suggests that a decrease in AMOC strength is associated with a predominant increase in sea-ice melt relative to sea-ice formation over the Atlantic Ocean surface. Figure \ref{F:Fig9} illustrates the trends in sea-ice formation (Figure \ref{F:Fig9a}a) and sea-ice melt (Figure \ref{F:Fig9b}b) rates for model years 1,200 to 1,700. In the North Atlantic, the sea-ice formation trend is mostly positive, contributing to a reduction in $M$. However, consistent with the data shown in Figure \ref{F:Fig7}, the dominant trend is an increase in the melting rate, which results in a net increase in $M$ across the North Atlantic. Approximately 77\% of the overall positive trend in the melt rate is attributed to increased basal, lateral, and top melt (not shown), while the remaining 23\% is due to heightened snow melt on top of the expanded North Atlantic sea-ice pack.

Given the increased net melt of sea ice over the North Atlantic region, it follows that these contributions must originate from outside the region.
The sea-ice pack is advected southward from regions north of 65$^\circ$N (see arrows in Figure~\ref{F:Fig9a}a). The coastal regions close to south(eastern) boundary of Greenland show negative melt-rate tendencies (lower temperatures due to AMOC weakening), resulting in sea-ice accumulation in the coastal areas. Then, via advection, the sea-ice pack enters the North Atlantic region and during winter time snow accumulates on top of the sea-ice pack. The following spring and summer, the sea-ice pack and snow melt contribute to a positive growth in $M$. This sequence of events highlights an AMOC-sea-ice feedback, which further destabilises the AMOC.

The last destabilising AMOC contribution comes from evaporation (Figure \ref{F:Fig9a}a) and precipitation (Figure \ref{F:Fig9b}b) changes.
The most notable changes in evaporation occur in the North Atlantic, where the enhanced sea-ice cover and sea surface cooling lead to a positive (i.e., less negative) evaporation tendency in response to an AMOC strength decline. The cooling effect also results in decreased precipitation over the North Atlantic. 
However, the evaporation trend dominates, leading to net freshwater into the North Atlantic due to changes in atmospheric fluxes. The opposite precipitation and evaporation responses over the North Atlantic region likely explain the weaker correlation between $P-E$ and $\Psi_{\mathrm{int}}$.
\begin{figure*}[htpb]
	\captionsetup{justification=centering}
	\centering
	\includegraphics[width=\textwidth]{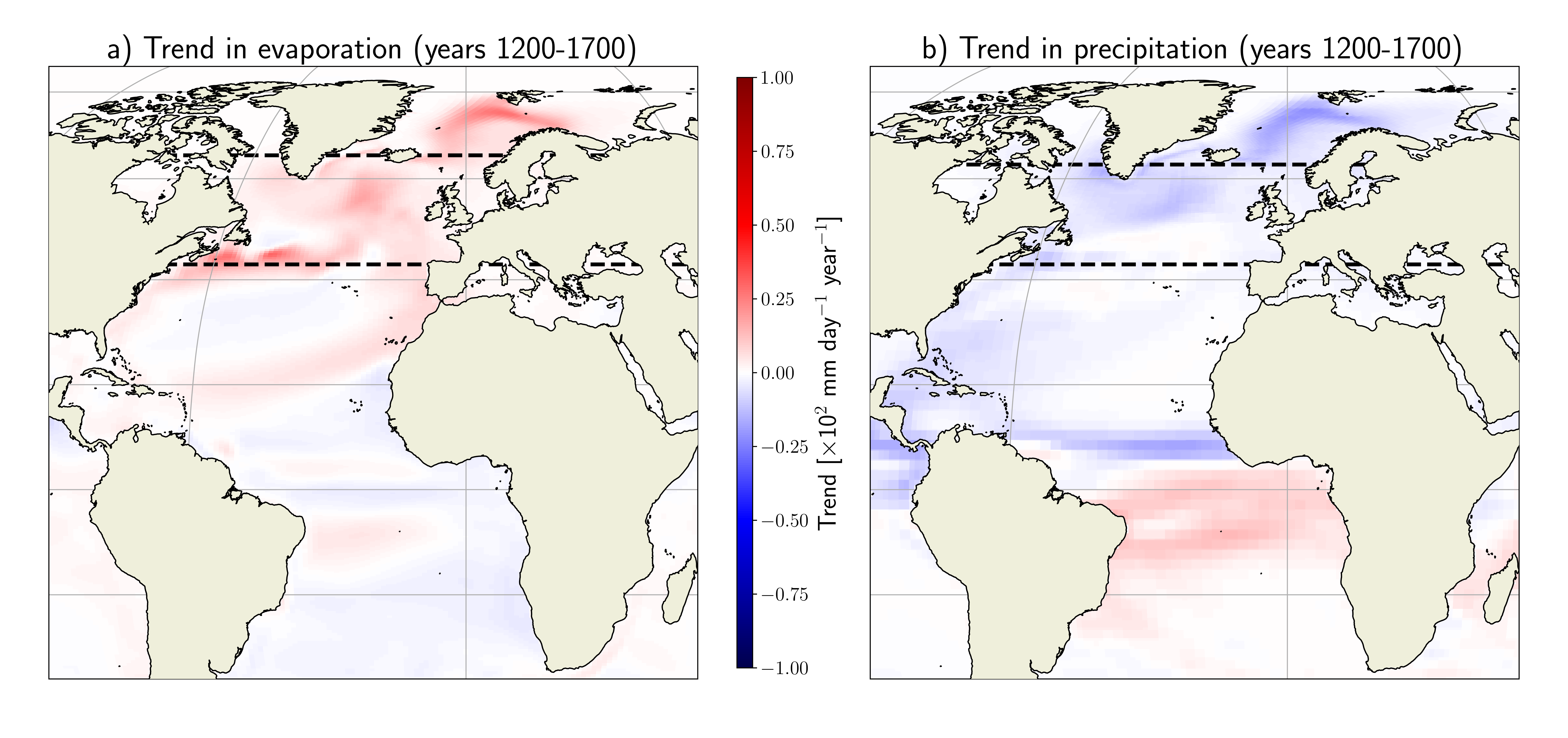}
	\caption{\em \small (a) Tendency of the yearly averaged evaporation rate for years 1000 -- 1500. (b) Tendency of the yearly averaged precipitation rate for years 1000-1500. Only significant ($p<0.05$) trends are shown. Black dashed curves mark the boundaries of the North-Atlantic.}
	\label{F:Fig10a}
	\label{F:Fig10b}
	\label{F:Fig10}
\end{figure*}
\subsection{Feedbacks from the Gyre circulation}
\begin{figure*}[htpb]
	\captionsetup{justification=centering}
	\centering
	\includegraphics[width=1\textwidth]{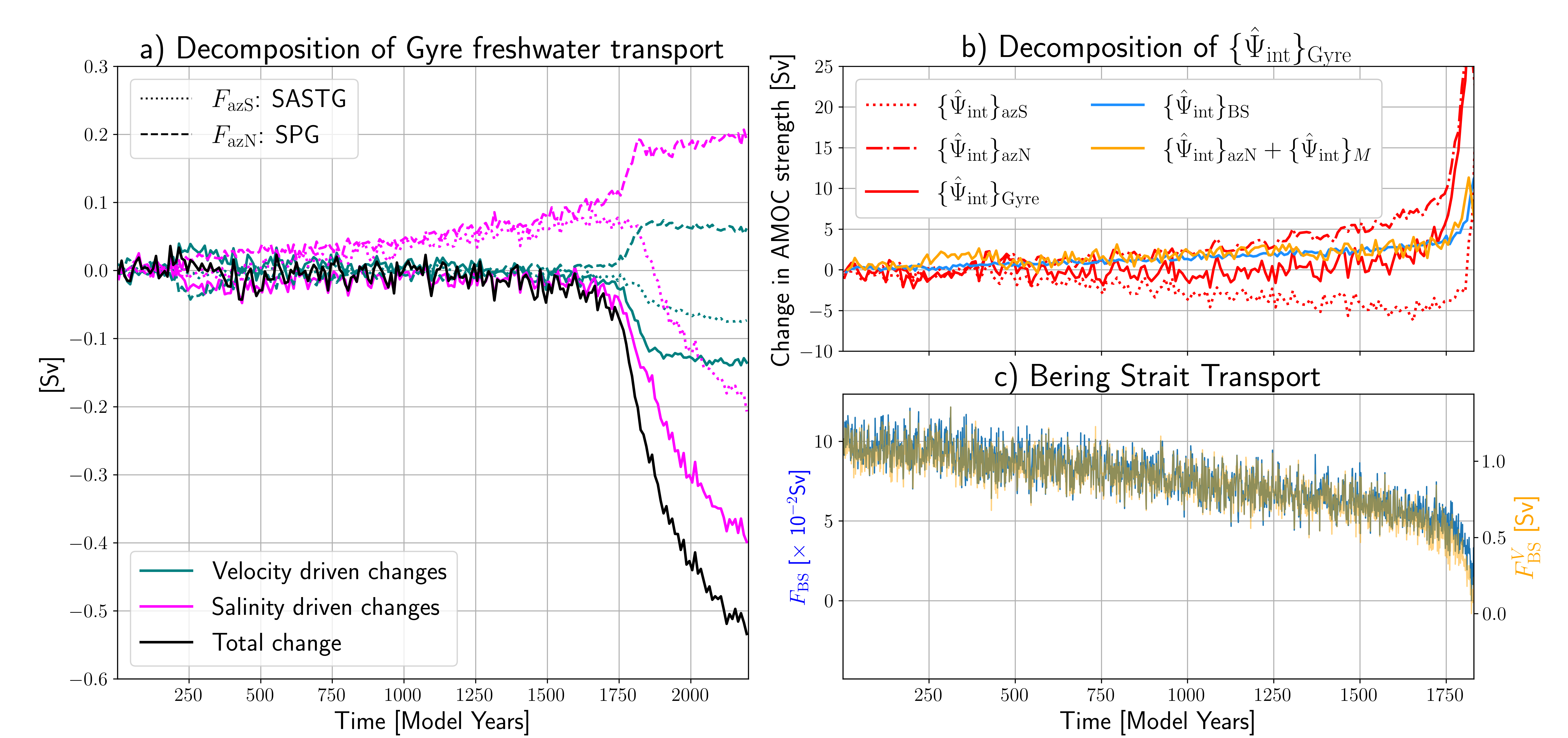}
	\caption{\em \small (a) Decomposition of change in Atlantic gyre freshwater convergence (thick black, $F_{\mathrm{azS}}-F_{\mathrm{azN}}$) into velocity (thick teal) and salinity (thick magenta). Dotted and dashed line represent decomposition for $F_{\mathrm{azS}}$ and $F_{\mathrm{azN}}$, respectively. (b): Contribution of $F_{\mathrm{azN}}$ (SPG, dash-dot red), $F_{\mathrm{azS}}$ (SASTG, dotted red) and Bering Strait freshwater transport (thick blue line) to change in AMOC strength. The Bering Strait contribution was argued to follow from $-\delta F_{\mathrm{azN}}+\delta M\approx \delta F_{\mathrm{BS}}$ (Figure \ref{F:Fig13}). (c): Bering Strait freshwater (left y-axis, thick blue line) and volume (right y-axis, thick orange line) transport.}
	\label{F:Fig11a}
	\label{F:Fig11b}
	\label{F:Fig11c}
	\label{F:Fig11}
\end{figure*}

Apart from the three destabilising AMOC contributions, the gyre contribution tends to stabilise the AMOC (Figure~\ref{F:Fig6a}a). 
The gyre circulation may modify through changes in the velocity or salinity field  \citep{drijfhout2011stability,weijer2019stability}. 
This velocity and salinity decomposition of the gyre freshwater convergence ($F_{\mathrm{azS}}-F_{\mathrm{azN}}$) and 
the separate contributions from $F_{\mathrm{azS}}$ (i.e., freshwater transport by the South Atlantic Subtropical Gyre (SASTG)) and $F_{\mathrm{azN}}$ (i.e., freshwater transport by the Subpolar Gyre (SPG)) are shown in Figure \ref{F:Fig11a}a. 
The gyre freshwater convergence changes are relatively small in the first 1,700 years. 
Closer to the tipping event, the freshwater convergence decreases and is primarily driven by changes in the salinity field.

\begin{figure*}[htpb]
	\captionsetup{justification=centering}
	\centering
	\includegraphics[width=1\textwidth]{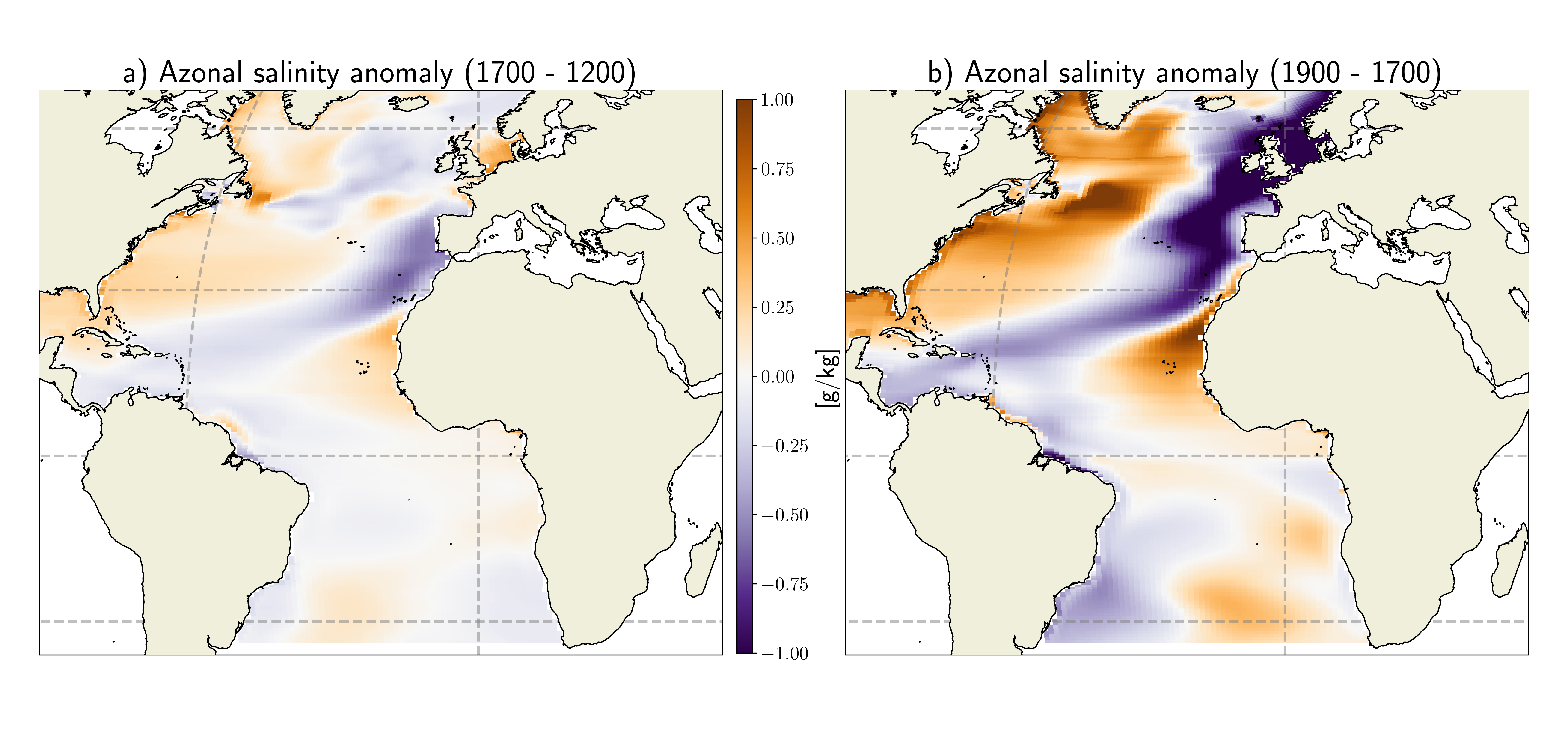}
	\caption{\em \small Vertically-averaged  ($\sim$ 0 -- 500 m) Atlantic Surface water (ASW)  azonal salinity difference, 
		with (a): model years 1,700 minus 1,200 and (b) 1,900 minus 1,700.}
	\label{F:Fig12a}
	\label{F:Fig12b}
	\label{F:Fig12}
\end{figure*}

Both $F_{\mathrm{azN}}$ and $F_{\mathrm{azS}}$ increase prior to tipping. This increasing trend is primarily driven by salinity changes, with a dominant salinity decrease at the eastern boundary in the North and South Atlantic (Figure \ref{F:Fig12a}a). Following equation (\ref{Process_Decomp}), the increase in $F_{\mathrm{azN}}$ works to stabilize the AMOC decline, while an increasing $F_{\mathrm{azS}}$ enhances the AMOC decline (Figure \ref{F:Fig11b}b). The contrasting of roles of the SPG and SASTG on AMOC stability before tipping agree with the findings of \cite{cimatoribus2014meridional}, and suggests that the SPG increases \citep{longworth2005ocean}, while the SASTG decreases the critical freshwater forcing required for AMOC collapse. Consequently, the total effect of the gyres on the critical freshwater forcing is reduced.

$F_{\mathrm{ovS}}$ reaches a minimum (model year 1,732, Figure \ref{F:Fig7a}a). This minimum arises as the velocity-induced responses become equally important as the salinity-induced responses, 
where eventually the velocity-induced responses start to dominate,  resulting in a $F_{\mathrm{ovS}}$ increase.
In other words, the AMOC carries less salinity into the Atlantic Ocean leading to salt accumulation in the southeast Atlantic Ocean (Figure \ref{F:Fig12b}b), as also described in \cite{zhu2020weakening}. 
This accumulation of salt results in a $F_{\mathrm{azS}}$ maximum around the same time as the $F_{\mathrm{ovS}}$ minimum.
Thereafter, the $F_{\mathrm{azS}}$ declines and contributes to an AMOC stabilisation of about 5~Sv (model years~1,600 -- 1,800).
This suggests that the SASTG tends to destabilize the northward overturning while also destabilizing the collapsed AMOC, consistent with results from earlier studies \citep{mecking2016stable,weijer2019stability}.

\begin{figure*}[htpb]
	\captionsetup{justification=centering}
	\centering
	\includegraphics[width=.5\textwidth]{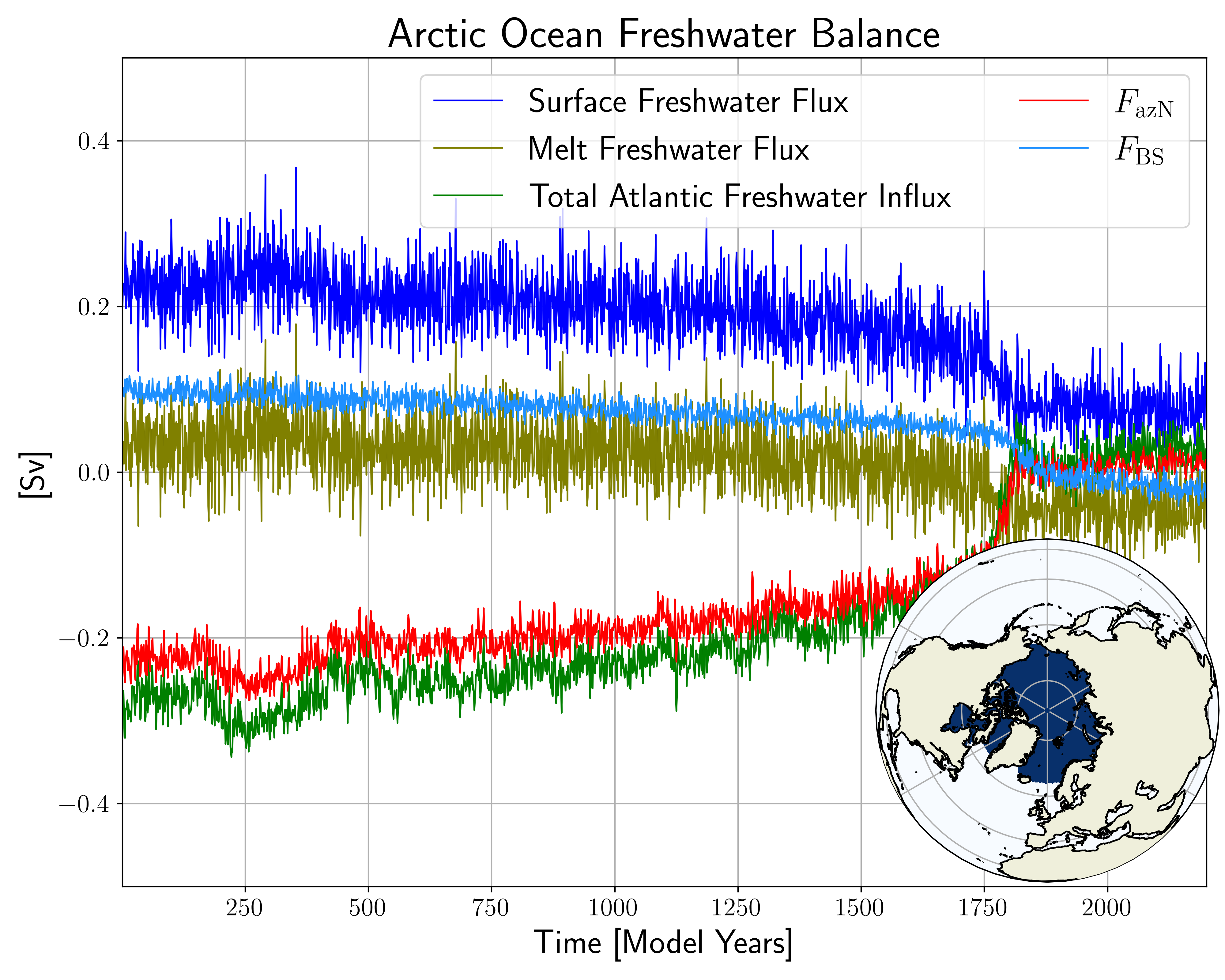}
	\caption{\em \small Arctic freshwater balance (equation (\ref{Balance})). The inset shows the Arctic region (blue).}
	\label{F:Fig13}
\end{figure*}

Figure \ref{F:Fig13} shows the Arctic Ocean freshwater balance. The salinity influx from the Atlantic Ocean is balanced by a surface freshwater flux and Bering Strait freshwater influx.
Under the increasing hosing freshwater flux forcing, the SPG reduces Atlantic salinity influx and this is compensated by a smaller surface freshwater flux and Bering Strait freshwater flux. 
The surface freshwater flux changes are primarly related to more sea-ice formation, where the sea-ice pack is advected out of the Artic Ocean and into the North Atlantic (cf. Figure \ref{F:Fig9}).
The dominant balance (Figure \ref{F:Fig13}) is then, $\delta F_{\mathrm{azN}}+\delta M_{\mathrm{ar}} \approx -\delta F_{\mathrm{BS}}$, where $M_{\mathrm{ar}}$ is the Arctic sea-ice melt freshwater flux which roughly satisfies $\delta M_{\mathrm{ar}} \approx \delta M$, with $M$ representing the Atlantic integrated melt surface freshwater flux (cf. Figure \ref{F:Fig8}). 
According to the framework described above (equation (\ref{Process_Decomp})), this balance can be equivalently written as $\delta\{\hat{\Psi}_{\mathrm{int}}\}_\mathrm{{azN}}+\delta\{\hat{\Psi}_{\mathrm{int}}\}_{\mathrm{Melt}}\approx \delta\{\hat{\Psi}_{\mathrm{int}}\}_{\mathrm{BS}}$ (Figure \ref{F:Fig11b}b). 
Hence, the stronger AMOC stabilization of the SPG, relative to the sea-ice melt feedback, can be attributed to the Bering Strait configuration \citep{hu2012role}. 
Specifically, in the CESM, where the Bering Strait is open, a weakened AMOC results in a decreased northward volume and freshwater transport through Bering Strait (Figure \ref{F:Fig11c}c).
This increases the Arctic surface salinity and thereby the salt transport of the southward flowing SPG branch, providing an additional input of salt into the North Atlantic in response to an AMOC weakening. 

\section{Summary and Discussion}\label{SD}

We investigated  the feedbacks involved in  AMOC tipping  within CESM under a quasi-equilibrium hosing \citep{van2023asymmetry}, aiming 
to resolve uncertainties surrounding the role of $F_{\mathrm{ovS}}$ as an AMOC stability indicator. Conceptual AMOC box models typically exhibit tipping behavior driven primarily by the salt-advection feedback. In such models, where the density contrast is determined by the Atlantic freshwater budget, $F_{\mathrm{ovS}}$ is considered a key indicator of AMOC stability \citep{rahmstorf1996freshwater}. However, modern climate models incorporate a broader range of AMOC feedback mechanisms, complicating the interpretation of $F_{\mathrm{ovS}}$ as a single stability metric. Furthermore, unlike box models \citep{cimatoribus2014meridional}, the link between AMOC stability and the Atlantic freshwater 
balance in comprehensive climate models is less clear.

Our approach relied on reconstructing the AMOC using the thermal wind balance, a framework strongly supported by  theory \citep{marotzke1997boundary,nikurashin2012theory,butler2016reconstructing}. This relation suggests that AMOC strength scales linearly with the twice-vertically-integrated density contrast between the North and South Atlantic. By linking the AMOC strength to the meridional density (or buoyancy) field,
we demonstrated that the dominant driver of AMOC weakening is the freshening of the North Atlantic Ocean.
The weaker AMOC transports less heat northward, which results in cooling of the North Atlantic Ocean.
This cooling stabilises the AMOC as it partially offsets (50\%) the AMOC weakening caused by  freshening \citep{van2024role, rahmstorf1996freshwater}. 
Additional adjustments in the South Atlantic further opposed the AMOC weakening, though their effect was smaller, accounting for a 20\% offset.

In analogy to the \cite{rahmstorf1996freshwater} box model, the vertical salinity contrast at 34$^\circ$S was found to be strongly correlated with the meridional density contrast. 
Employing thermal wind balance, this correlation allowed us to deduce that AMOC stability can be analyzed from a steady-state Atlantic integrated freshwater balance.
Using this framework, we quantified the contributions of various feedback mechanisms that destabilize the northward overturning circulation. 
The imposed hosing directly weakens the AMOC, but here we focus on the feedbacks that amplify the imposed hosing.
The overturning salt-advection feedback is the dominant destabilising mechanism and is responsible for the AMOC collapse. 
The salt-advection feedback response is related to the $F_{\mathrm{ovS}}$ sign. 
When $F_{\mathrm{ovS}} < 0$, the salt-advection destabilizes the AMOC and the opposite is true for $F_{\mathrm{ovS}} > 0$.
Furthermore, the strength of the destabilizing salt-advection feedback increases disproportionately with decreasing $F_{\mathrm{ovS}}$. 
Our findings are therefore consistent with previous studies \citep{huisman2010indicator,sijp2012characterising}, without relying on the assumption that AMOC stability relates to the Atlantic freshwater content \citep{van2024physics}.

In addition to the overturning salt-advection feedback, other destabilizing mechanisms were identified, such as enhanced sea-ice melt 
and increased atmospheric freshwater flux in response to AMOC weakening. The gyre circulation played a key role in stabilizing the AMOC during hosing, with the subpolar gyre in the North Atlantic providing the primary stabilization. While the subtropical gyre in the South Atlantic contributes to AMOC 
destabilization, the North Atlantic gyre’s stabilizing effect is dominant. This contrasting influence of the two gyres aligns with box 
model results from \cite{cimatoribus2014meridional}. The strong stabilization by the North Atlantic subpolar gyre may be linked to the open Bering Strait configuration in CESM. In response to an AMOC decline, this configuration allows additional salt to enter the Arctic Sea, which is then transported to the outcropping region by the gyre's southward branch, as suggested by \cite{hu2012role}. 
However, we acknowledge that the CESM version analyzed here lacks eddy-resolving dynamics, which have been shown to significantly affect the Atlantic freshwater balance and AMOC stability \citep{mecking2016stable,toom2014response,juling2020atlantic}. The contribution of eddies to AMOC stability remains a subject for future investigation.


Shown in Figure \ref{F:Fig4a}a, the haline ($\{\hat{\Psi}_{\mathrm{int}}\}_{\langle\Delta_y S\rangle}$) and thermally driven ($\{\hat{\Psi}_{\mathrm{int}}\}_{\langle\Delta_y T\rangle}$) AMOC changes display strong opposing variability on a multi-centennial timescale. 
The origin of this multi-centennial variability has been attributed to the salt-advection feedback \citep{li2022theory} and its amplitude is damped by the thermal advective overturning feedback \citep{yang2024theory}. 
Figure \ref{F:Fig4a}a shows that this variability originates from vertically-averaged density variations in the North Atlantic, consistent with recent work \citep{yang2024north,cheng2018can}.
Figure \ref{F:Fig5}d showed that this multi-centennial variability is not captured in the AMOC reconstruction at 34$^\circ$S ($\{\hat{\Psi}_{\mathrm{int}}\}_{\Delta_v S}$). 
This suggests that buoyancy anomalies driving the variability remain localized to the North Atlantic. Therefore, they operate independently of the Atlantic freshwater balance, or $F_{\mathrm{ovS}}$. 
This explains why recent studies \citep{haines2022variability,cheng2018can,mignac2019decoupled} found no significant influence of South Atlantic freshwater transport anomalies on AMOC strength under statistically stationary conditions.
However, the 34$^\circ$S AMOC reconstruction does capture the persistent weakening of the AMOC under increasing hosing strength. This implies that adjustments toward a new equilibrium under altered forcing, such as hosing, depend on the Atlantic freshwater budget, particularly $F_{\mathrm{ovS}}$. 
Specifically, equations (\ref{Process_Decomp}) and (\ref{SAF_Sol}) indicate that a more negative $F_{\mathrm{ovS}}$ requires a larger decline in AMOC strength for each unit of North Atlantic hosing increase.

Our study provides further evidence that $F_{\mathrm{ovS}}$ measures the stability of a (statistically) stationary 
AMOC state and hence provides a measure of  the distance from the critical freshwater forcing threshold 
\citep{rahmstorf1996freshwater}.  While an integrated freshwater budget may not be ideal for assessing stability 
under rapid transient forcing (e.g., climate change), the initial $F_{\mathrm{ovS}}$ value still provides a useful 
estimate of the AMOC response to buoyancy  anomalies \citep{liu2017overlooked,vanwesten2024substantialrisk21stcentury}. 
Since most CMIP5 and CMIP6 models exhibit a positive $F_{\mathrm{ovS}}$ \citep{mecking2017effect,van2024persistent}, 
the AMOC in their background climate  states is overly stable, leading to an underestimation of the risk of AMOC tipping.
It is therefore crucial that the climate models participating in CMIP7 reduce their $F_{\mathrm{ovS}}$ biases to better 
capture AMOC responses to climate change. 

\section*{Acknowledgments} 
E.Y.P.V., R.M.v.W. and H.A.D. are funded by the European Research Council through the ERC-AdG 
project TAOC (project 101055096, PI: Dijkstra). The authors gratefully acknowledge Michael Kliphuis (IMAU, UU) 
for performing the model simulations presented in this study. The CESM model simulations and the analysis of 
all the model output was conducted on the Dutch National  Supercomputer Snellius within NWO-SURF project 
2024.013.

\section*{Availability Statement}
The analysis scripts will be made available on Zenodo upon
publication.

\section*{Appendix A: Derivation of Equation (\ref{SAF_Sol})}
\renewcommand{\theequation}{A.\arabic{equation}}
\renewcommand{\thefigure}{A.\arabic{figure}}
\setcounter{equation}{0}
\setcounter{figure}{0}
Figure \ref{F:Fig3} shows that we can approximate the AMOC strength  as:
\begin{equation}
	\Psi_{\mathrm{int}} \approx \Psi_{\mathrm{int}}(0) + \{\hat{\Psi}_{\mathrm{int}}\}_{\langle\Delta_y \rho\rangle},
\end{equation}
From Figure \ref{F:Fig5a}, it follows that $\{\hat{\Psi}_{\mathrm{int}}\}_{\langle\Delta_y T\rangle}$ scales linearly 
with $-\{\hat{\Psi}_{\mathrm{int}}\}_{\langle\Delta_y S\rangle}$. This makes sense  given that, prior to AMOC tipping, 
changes in North Atlantic temperature result from the weakening of the overturning circulation, which in turn 
is linked to the North Atlantic freshening. Therefore, we can express the overturning streamfunction as:
\begin{equation}
	\Psi_{\mathrm{int}}(t) \approx \Psi_{\mathrm{int}}(0) + (1 - c_1)\{\hat{\Psi}_{\mathrm{int}}\}_{\langle\Delta_y S\rangle}(t),
	\label{Delta_Psi_Approx}
\end{equation}
where the constant $c_1$ quantifies the efficiency of the thermal advective feedback in offsetting the 
salinity-driven AMOC decrease. We assume that $c_1$ is independent of time, which is supported by 
statistical analysis (Figure \ref{F:FigA1}).

The linear relation between $\langle \Delta_y S \rangle$ and $\Delta_v S$ can now be used to express:
\begin{equation}
	\{\hat{\Psi}_{\mathrm{int}}\}_{\langle\Delta_y S\rangle}(t) = -\frac{gCD^2\beta \mu}{f}\left(\Delta_v S(t) - \Delta_v S(0)\right).
	\label{DeltaS_Approx}
\end{equation}
Combining equations (\ref{Delta_Psi_Approx}) and (\ref{DeltaS_Approx}), we can rewrite the expression for $\Psi_{\mathrm{int}}$ as:
\begin{equation}
	\Psi_{\mathrm{int}}(t) \approx \frac{gCD^2\beta \mu}{f}\left(\underbrace{\frac{\Psi_{\mathrm{int}}(0) f}{gCD^2\beta \mu} + (1-c_1)\Delta_v S(0)}_a - (1 - c_1)\Delta_v S(t)\right).
	\label{Approximation_DeltaS}
\end{equation}
Assuming that the time instances $t_1$ and $t_2$ are infinitesimally close, $\delta \Psi_{\mathrm{int}}$ becomes 
$d\Psi_{\mathrm{int}}$. Inserting equation (\ref{Approximation_DeltaS}) into the definition of the salt-advection feedback contribution (equation (\ref{Process_Decomp})) and integrating over $\Delta_v S$ we express:
\begin{equation}
	\{\hat{\Psi}_{\mathrm{int}}\}_{\mathrm{Saf}} = -\frac{g\mu C \beta D^2 (1 - c_1)}{f} \left( \int_{\Delta_v S(0)}^{\Delta_v S(t)} \frac{\Delta_v S'}{a - (1 - c_1)\Delta_v S'} d(\Delta_v S)' \right).
\end{equation}
It then follows:
\begin{equation}
	\{\hat{\Psi}_{\mathrm{int}}\}_{\mathrm{Saf}} \approx \frac{g\mu C \beta D^2}{f(1 - c_1)} \left( (1 - c_1)(\Delta_v S(t) - \Delta_v S(0)) + a \log\left(\frac{a-(1-c_1)\Delta_v S(t)}{a-(1-c_1)\Delta_v S(0)}\right) \right).
\end{equation}

\begin{figure*}[htpb]
	\captionsetup{justification=centering}
	\centering
	\includegraphics[width=.5\textwidth]{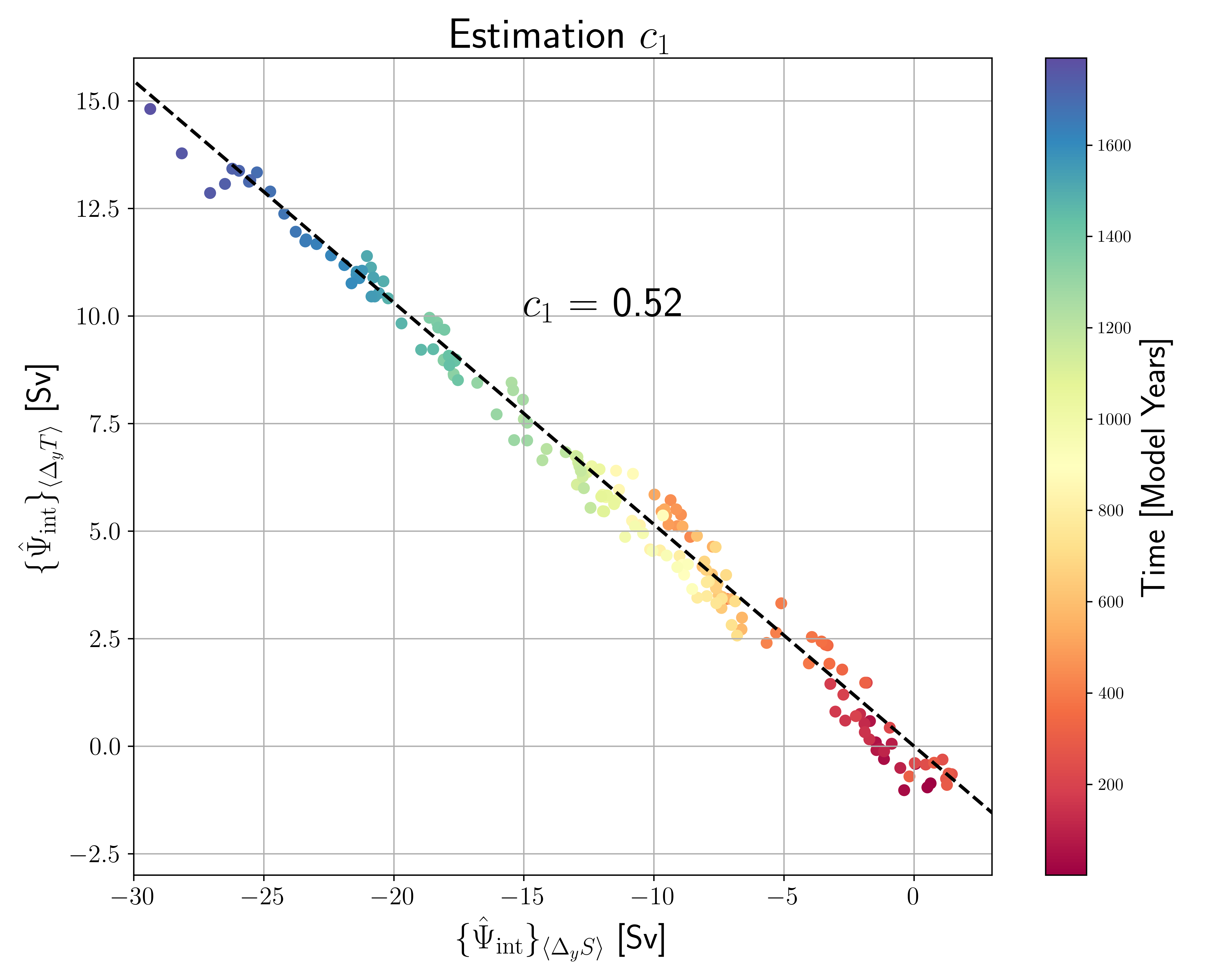}
	\caption{\em \small Relation between $\{\hat{\Psi}_{\mathrm{int}}\}_{\langle \Delta_y S\rangle}$ and $\{\hat{\Psi}_{\mathrm{int}}\}_{\langle \Delta_y T\rangle}$, where each dot represents a 10-year average. Slope of relation determines value of $-c_1$.}
	\label{F:FigA1}
\end{figure*}

\bibliographystyle{ametsocV6}
\bibliography{bibliography.bib}
\end{document}